\documentclass[preprint,fleqn,5p,numbers,sort,compress]{elsarticle}
\usepackage{caption} 
\usepackage{amssymb, amsmath, amsfonts, latexsym}
\journal{Journal of \LaTeX\ Templates}
\usepackage{hyperref}
\hypersetup{pdftitle = The title of my PDF, pdfauthor = My name, pdfsubject= The subject, pdfkeywords = keyword1 keyword2 keyword3}
\hypersetup{colorlinks = true, linkcolor = blue, anchorcolor = red, citecolor = blue, filecolor = red, pagecolor = red, urlcolor = blue}

\begin{document}

\begin{frontmatter}

\title{Unpredictable basin boundaries in restricted six-body problem with square configuration}

\author[vk]{Vinay Kumar\corref{cor1}}


\author[ji]{M. Javed Idrisi\corref{cor2}}
\author[su]{M. Shahbaz Ullah\corref{cor3}}
\ead{mdshahbazbgp@gmail.com}

\cortext[cor3]{Corresponding author}

\address[vk]{Department of Mathematics, Zakir Husain Delhi College, University of Delhi, New Delhi 110002, India}
\address[ji]{Department of Mathematics, College of Natural and Computational Science, Mizan-Tepi University, Tepi Campus, Ethiopia}
\address[su]{Department of Mathematics, Zakir Husain Delhi College, University of Delhi, New Delhi 110002, India}


\begin{abstract}
The present work deals with the recently introduced restricted six body-problem with square configuration.  It is determined that the total number of libration points are twelve and twenty for the mass parameter $0< \mu < 0.25$. The multivariate form of Newton-Raphson scheme is used to discuss the basin of attraction. Different aspects of the basin of attraction are investigated and explained in detail. The complex combination of the different basins is found along the boundaries. The concept of basin entropy is used to unveil the nature of the boundaries. For $\mu = 0.22$ and $0.23$, the basin of attraction is unpredictable throughout. It is observed that for all values of the mass parameter $\mu$, the basin boundaries are highly unpredictable.  Further, We have investigated the presence of Wada basin boundary in the BoA. 
\end{abstract}

\begin{keyword}
\texttt{Restricted six-body problem (R6BP), Newton-Raphson basin of attraction (N-R BoA), Fractal basin boundary, Wada basin}
\MSC[2010] 70H07\sep 37N05
\end{keyword}

\end{frontmatter}


\section{Introduction}
In the field of \emph{Celestial Mechanics}, the N bodies problem has very significant contributions. It has numerous applications in the field of galactic dynamics, the motion of planetary objects. Many articles are available for the N-bodies for N=3, 4 and 5 (\cite{Arribas16}, \cite{Balta2011},   \cite{Celli2007},  \cite{Kalv1999}, \cite{Marchesin2017}, \cite{Michalo1981a},  \cite{Papad2013},   \cite{Shoaib2011} and \cite{Suraj2019c}). Recently, we have introduced the general extension of this N-body problem known as restricted six-body problem with square configuration (\cite{Idrisi2020}). Therefore a lot of works have to be done in this model. The restricted problem of six-bodies is to study the motion of the test particle under the gravitational field of four primaries placed on the vertices of the square while one primary is placed on the centre of mass of the system. We have considered the mass ratio $\mu$ as the only parameter.

In general, for the problem of N-bodies ($N>4$), there is no specific method to determine the number of libration points. Therefore, we usually find it using numerical methods. Now there are various numerical methods available to find out the libration points (or roots) of these dynamical systems. Among them, the Newton-Raphson (N-R) scheme is very well known and established method to determine the roots of nonlinear dynamical systems. In our case, we need the multivariate form of the N-R scheme. While going through recent articles on the applications of N-R method (\cite{Dubeibe2020},  \cite{Suraj2019a} and \cite{Suraj2019}), we note that some initial conditions converge very quickly, some of them need more number of iterations, some of them even do not converge to any of the libration points. Thus, the study of the convergence of initial conditions is also a crucial aspect of the investigation. Also, it is essential to note that initial conditions lying along the boundary need a higher number of iterations. Therefore, the detailed study of the basin of attraction (BoA) in R6BP is also one of the critical aspects. We can see some articles related to it in the work of (\cite{sprott2015}, \cite{Suraj2019a} and \cite{Suraj2019}, (including their references))

The applications of the N-R scheme to the restricted problem of N bodies can be found in the work of (\cite{Suraj2019a}, \cite{Suraj2019b}, \cite{Suraj2020} and \cite{Suraj2019}, ). Based on that,  we have investigated the BoA in R6BP using the multivariate form N-R scheme. In many cases, the BoA is found to be smooth except few.  However, when the basins are not smooth, then we search for the degree of the unpredictability in BoA and along boundaries of BoA. To measure this, we use the concept of basin entropy introduced recently (\cite{Alvar16}). The configuration plane $(x,y)$ can be divided into two parts; one is a fractal region and other is a non-fractal region (based on $\log2$ graph shown in Figs. 7). We can decide based on the values of the basin entropy ($S_{b}$) and the boundary basin entropy ($S_{bb}$) obtained using the algorithm explained in (\cite{Alvar16}). When there is a coexistence of two or more attractors, there is a possibility for the occurrence of an important property called Wada. The concept and algorithm to show Wada basin boundary can be seen in the work of (\cite{Aguirre2009},  \cite{Sanjuan2001}, \cite{Sanjuan2002}, \cite{Alvar17}, \cite{Daza2015}, \cite{Alvar17a}, \cite{Sanjuan2018} and \cite{Sanjuan2013}). We have also investigated the existence of Wada basin boundary in the R6BP.

Thus, in the present work, we have considered R6BP for investigation. To explore the BoA, we consider the permissible value of the parameter $\mu$ $\in$ (0, 0.25) and the N-R scheme. One of the crucial aspects is to explore the existence of an unpredictable region in BoA and along the boundaries. Further, the possibility of Wada basin boundary is also examined. 

We have organised the present work as follows:
The configuration and equations of motion of R6BP are explained in Section 2. The distribution of the potential function around the libration points is discussed in its subsection. In Section 3, we have mentioned the concept and algorithm used to find out BoA, basin entropy and boundary basin entropy. Results based on these concepts and the presence of wada basin boundaries are discussed in detail in Section 4. Concluding remarks based on numerical simulations and results are given in Section 5.

\section{Configuration of the restricted six-body problem}

The four particles $P_{i}$ of equal masses $m_{i}$, $i = 1,2,3,4$, respectively, are placed at the vertices of square, revolving with angular velocity $\omega$ in circular orbit about their common center of mass $O$ and the fifth particle $P_{0}$ of mass $m_{0}$ rests at the center. Let $OP_{i}=a$ be the distances of primaries from the center of mass, and an infinitesimal mass $\acute{m}$ is moving under the gravitational field of $m_{i}$ in the $xy$-plane. In such a system the motion of the infinitesimal is two dimensional. In the inertial frame of reference, the orbit is located in the $Oxy$ plane and its center as an origin. Since, the four primaries $P_{1}$, $P_{2}$, $P_{3}$, $P_{4}$ form a square and moving in circular orbit around their common center of mass; the particles $P_{1}$, $P_{3}$ and $P_{2}$, $P_{4}$ always lie on $x-axis$ and $y-axis$, respectively. The particle attracts each other under the Newtonian law of gravitation and forms a symmetric square-configuration with respect to the origin at any instant of time (Fig. 1).
\begin{figure}
	\centering
	\includegraphics[width=1.\columnwidth]{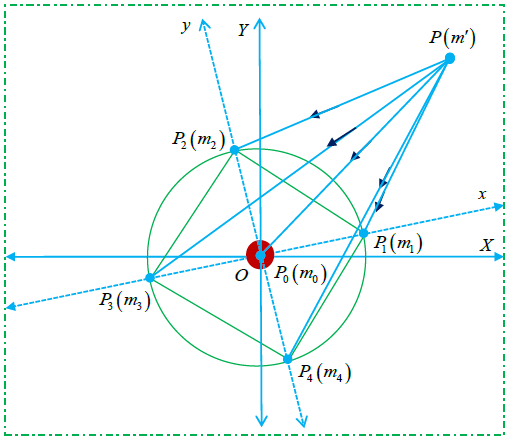}
	\caption{R6BP with square configuration }\label{hggyug}
\end{figure}

\begin{figure*}[htb!]
	\begin{tabular}{cc}
		 \includegraphics[scale=.41]{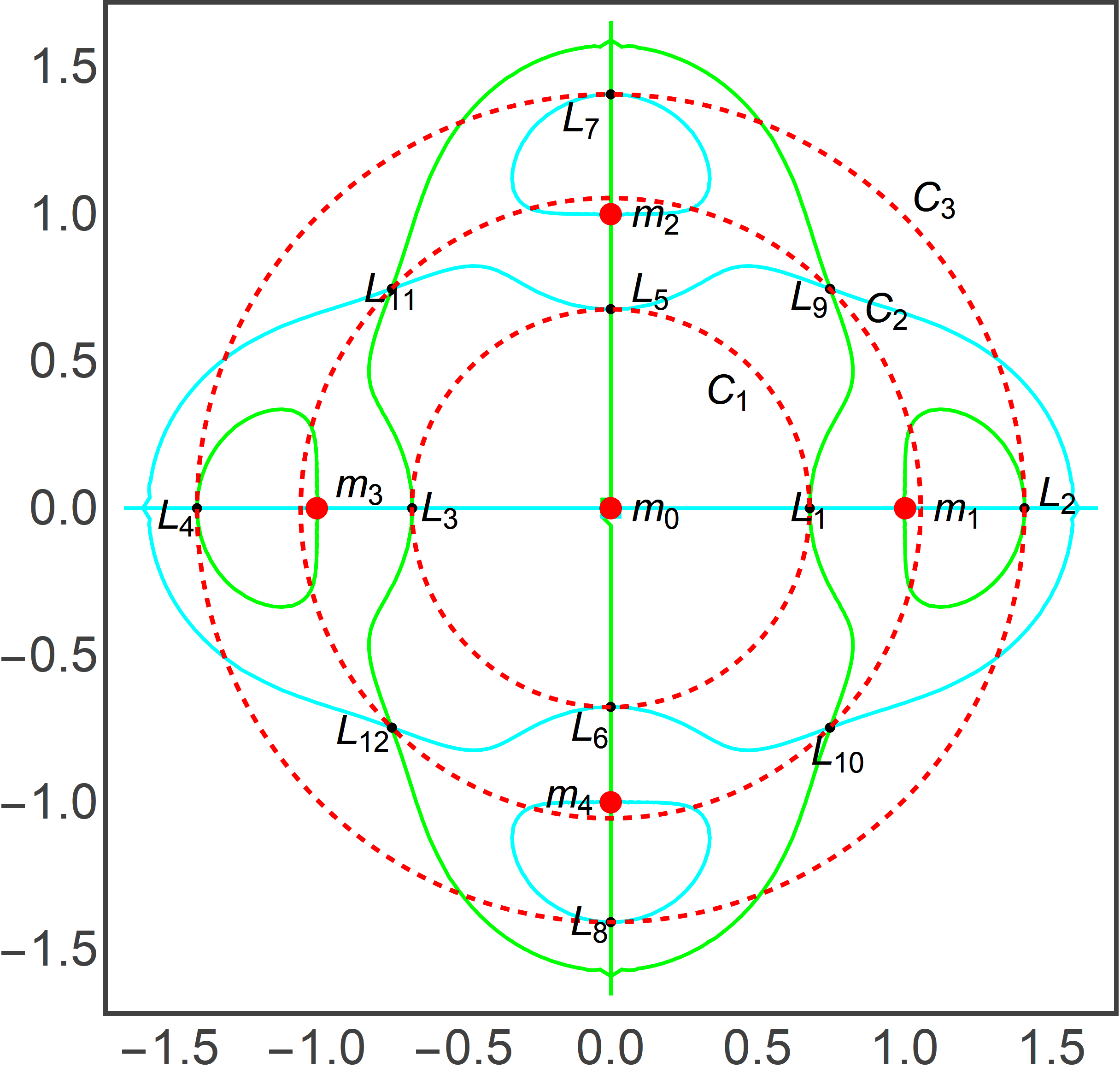}~~~~~~~~~~~~~~~~&\includegraphics[scale=.33]{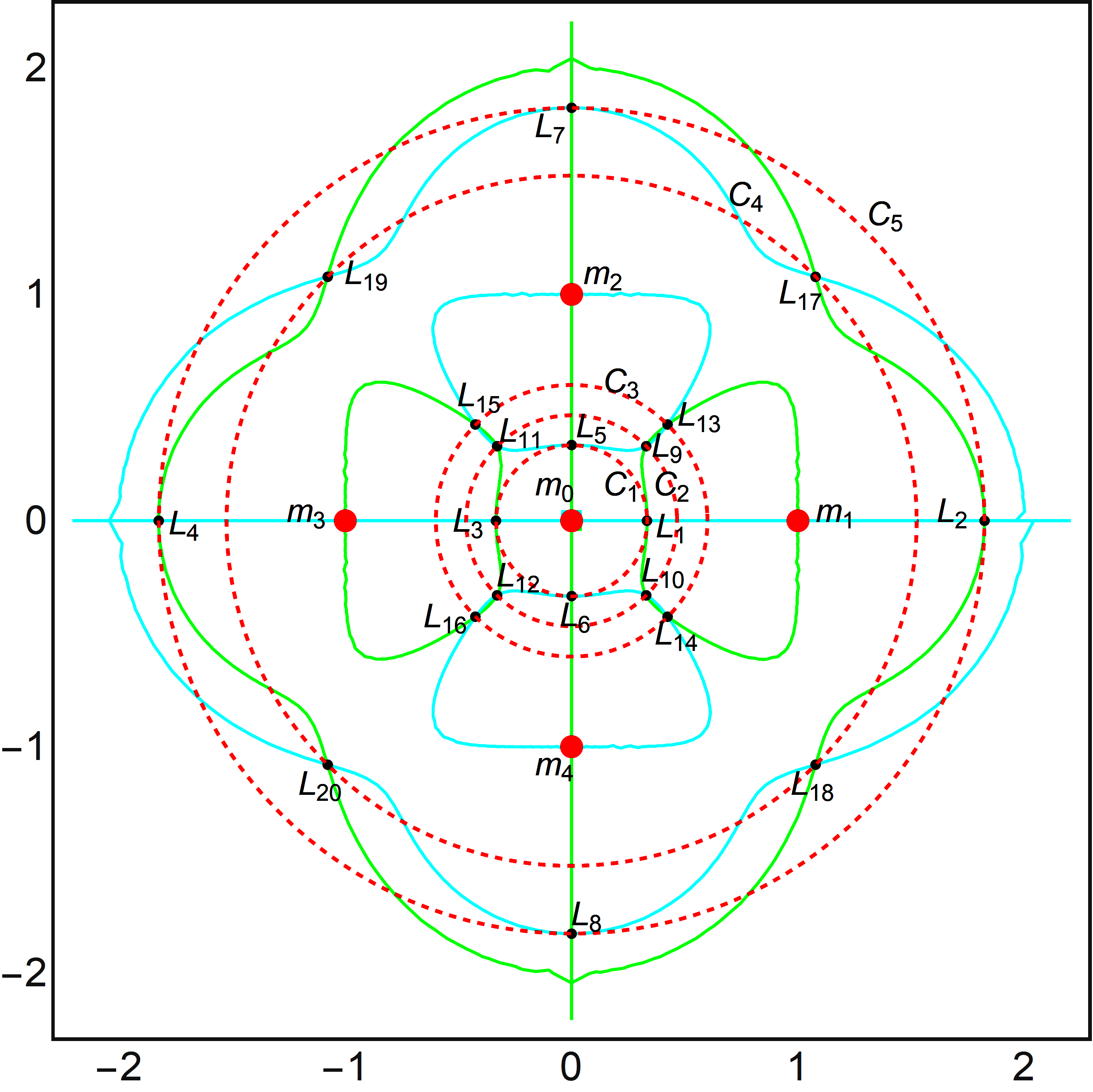}\\
		(a)  $\mu =$ 0.11& (b) $\mu =$ 0.24
	\end{tabular}
	\caption{The black dots indicate the positions of the libration points through the intersections of $\Omega_{x} = 0$ (green) and $ \Omega_{y} = 0$ (cyan), the red dots denote the position of primaries. (a) twelve libration points (b) twenty libration points are shown.}
\end{figure*}

\subsection{Equations of motion}
Let us consider the line joining $P_{1}$ and $P_{3}$ is taken as $X-axis$, $O$ their center of mass as origin, the line passing through $O$ and perpendicular to $OX$ lying in the plane of motion of $m_{i}'s$ as the $Y-axis$. We consider a synodic system of coordinates $O(xyz)$; initially coincident with the inertial system $O(XYZ)$, rotating with the angular velocity $\omega=n k$ about $Z-axis$. The distances of $\acute{m}$ from $m_{i}$  and $O$ are $r_{i}$  and $r_{0}$ respectively.

The equations of motion of the particle $P(x, y)$ having infinitesimal mass $\acute{m}<<1$ in the synodic coordinate system and dimensionless variables are (\cite{Idrisi2020}):
\begin{eqnarray*}
	\ddot{x}-2n\dot{y}=\Omega_{x},\\
	\ddot{y}+2n\dot{x}=\Omega_{y},
\end{eqnarray*}
where the potential function $\Omega$ can be expressed as
\begin{equation}
\Omega=\frac{n^2}{2}(x^2+y^2)+\frac{1-4\mu}{r_{0}}+\sum_{i=1}^{4}\frac{\mu}{r_{i}},
\end{equation}
$n$ is the mean-motion of the primaries defined as $n^2=1-c\mu; \mu=m/M$ is the mass parameter $0<\mu<1/4$; M is the sum of masses of all primaries taken as unity and $c=(15\sqrt{2}-4)/4\sqrt{2}$,                                                                                                       \begin{eqnarray*}
	r_{0}^2=x^2+y^2, r_{1}^2=(x-1)^2+y^2, r_{2}^2=x^2+(y-1)^2,
\end{eqnarray*}
\begin{eqnarray*}
	r_{3}^2=(x+1)^2+y^2, r_{4}^2=x^2+(y+1)^2.
\end{eqnarray*}
The Jacobi integral associated to the problem is
\begin{equation}
v^2=2 \Omega-\acute{C},
\end{equation}
where $v$ is the velocity of infinitesimal mass $\acute{m}<<1$ and $\acute{C}$ is Jacobi constant.

\subsection{Lower limit of potential function $\Omega$ and Jacobian constants}
The libration points are the solution of the Eqns. $\Omega_{x}=0$ and $\Omega_{y}=0, i.e.,$
\begin{eqnarray}
\ & \Omega_{x}=n^2 x -\nonumber\\
\ & \left\{  \frac{(1-4\mu)x}{r_{0}^3}+\frac{\mu (x-1)}{r_{1}^3}+\frac{\mu x}{r_{2}^3}+
\frac{\mu (x+1)}{r_{3}^3}+\frac{\mu x}{r_{4}^3}
\right\}=0
\end{eqnarray}
\begin{eqnarray}
\ & \Omega_{y}=n^2 y -\nonumber\\
\ & \left\{  \frac{(1-4\mu)y}{r_{0}^3}+\frac{\mu y}{r_{1}^3}+\frac{\mu (y-1)}{r_{2}^3}+
\frac{\mu y}{r_{3}^3}+\frac{\mu (y+1)}{r_{4}^3}
\right\}=0
\end{eqnarray}
Solving Eqns. (3) and (4), their exit 12 libration points out of which four libration points are collinear and eight are non-collinear and all the libration points lie on the concentric circles $C_{1}$, $C_{2}$ and $C_{3}$ centered at origin (\cite{Idrisi2020}). The libration points $L_{1}$, $L_{3}$, $L_{5}$ and $L_{7}$ are lying on circle $C_{1}$; $L_{9}$, $L_{10}$, $L_{11}$ and $L_{12}$ on $C_{2}$ and $L_{2}$, $L_{4}$, $L_{6}$ and $L_{8}$ on $C_{3}$. This is also observed that the eight libration points are on the axes and four are off the axes, $i.e.$, $L_{i} (i = 1, 2, 3, 4)$ are on $x-axis$, $L_{j} (j = 5, 6, 7, 8)$ on $y-axis$ and rest are off the axes (Fig. 2). The libration points are intersection of the curves $\Omega_{x}(x,y) =0$ and $\Omega_{y}(x,y)=0$. We have plotted the contour curves of $\Omega_{x}(x,y)$ and $\Omega_{y}(x,y)$ in Figs. 2(a) and 2(b) for $\mu=0.11$ and $\mu=0.24$, respectively. We observed that there exist twelve and twenty libration points for the parameter $\mu$ in (0, 0.25).

The libration points on x-axis ($L_{1}$, $L_{2}$, $L_{3}$, $L_{4}$) and $y-axis$ ($L_{5}$, $L_{6}$, $L_{7}$, $L_{8}$) are the solution of Eqns. (5) and (6), respectively.
\begin{eqnarray}
\ & f(x)=n^2 x-\nonumber\\
\ & \frac{(1-4\mu)x}{\mid x \mid^3}-\frac{\mu(x-1)}{\mid x-1 \mid^3}-\frac{2\mu x}{\mid x^2+1 \mid^3}-\frac{\mu(x+1)}{\mid x+1 \mid^3} = 0,
\end{eqnarray}
\begin{eqnarray}
\ & f(y)=n^2 y-\nonumber\\
\ & \frac{(1-4\mu)y}{\mid y \mid^3}-\frac{\mu(y-1)}{\mid y-1 \mid^3}-\frac{2\mu y}{\mid y^2+1 \mid^3}-\frac{\mu(y+1)}{\mid y+1 \mid^3} = 0.
\end{eqnarray}
The coordinates of libration points in $xy-plane$ ($L_{9}$, $L_{10}$, $L_{11}$, $L_{12}$) are $\left\{ (1+\delta)/\sqrt{2}, (1+\delta)/\sqrt{2} \right\}$,\\
$\left\{ -(1+\delta)/\sqrt{2}, -(1+\delta)/\sqrt{2} \right\}$,\\
$\left\{ -(1+\delta)/\sqrt{2}, (1+\delta)/\sqrt{2} \right\}$, and\\
$\left\{ (1+\delta)/\sqrt{2}, -(1+\delta)/\sqrt{2} \right\}$ respectively,\\
where
\begin{equation}
\delta=0.296884\mu+1.73206\mu^{2}.
\end{equation}
For detail please see:
\emph{Central-body square configuration of restricted six-body problem (\cite{Idrisi2020}).}
So, at the libration points the potential function is given as,
\begin{equation}
\Omega=\frac{\acute{C}}{2}=\frac{n^2}{2}(x_{0}^2+y_{0}^2)+\frac{1-4\mu}{\acute{r_{0}}}+\sum_{i=1}^{4}\frac{\mu}{\acute{r_{i}}},
\end{equation}
where
\begin{eqnarray*}
	\acute{r_{0}^2}=x_{0}^2+y_{0}^2, \acute{r_{1}^2}=(x_{0}-1)^2+y_{0}^2, \acute{r_{2}^2}=x_{0}^2+(y_{0}-1)^2,
\end{eqnarray*}
\begin{eqnarray*}
	\acute{r_{3}^2}=(x_{0}+1)^2+y_{0}^2, \acute{r_{4}^2}=x_{0}^2+(y_{0}+1)^2,
\end{eqnarray*}
and $x_{0}$, $y_{0}$ are the coordinates of libration points $L_{j}$ $(j = 1,..., 12)$ in the orbital plane of primaries.
\begin{figure}
	\centering
	\includegraphics[width=1.\columnwidth]{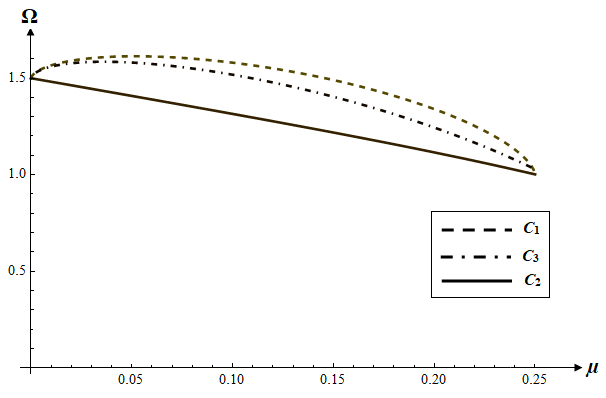}
	\caption{Graphics of $\Omega$ as a function of $\mu$; Solid curve corresponds to the libration points on circle $C_{2}$; dash-point-dash curve to the libration points on circle $C_{3}$; dash-dash to the libration points on circle $C_{1}$}\label{hggyug}
\end{figure}

From Fig. 3, it is observed that the potential function $\Omega$ has the maximum value for the libration points on circle $C_{1}$, while the minimum value for the libration points on $C_{2}$. For the libration points on the circles $C_{1}$ and $C_{3}$, as $\mu$ increases, $\Omega$ increases in the interval $0 < \mu < 0.05$ and then starts decreasing in $0.05 < \mu < 0.25$. For the libration points on the circle $C_{2}$, as $\mu$ increases, $\Omega$ decreases uniformly (Tables 1 and 2). Thus, $\Omega_{min}$ is defined as,\\
$\Omega_{C_{2}} = \Omega_{min} = 1.5 - 1.82593\mu - 0.13221\mu^{2} - O(\mu^{3})$.

It may be noticed that as $\mu \rightarrow 0$, $\Omega_{min} \rightarrow 1.5$ which is similar to the classical case of restricted three-body problem (Szebehely, 1967) and as $\mu \rightarrow 0.25$, $\Omega_{min} \rightarrow 1$. Thus, it is concluded that the infinitesimal mass $\acute{m}$ requires minimum energy depending upon the mass parameter $\mu$ to reach in the vicinity of libration points lying on circle $C_{2}$ while it requires maximum energy to reach in the vicinity of libration points lying on circle $C_{1}$.

\begin{table}
	\centering
	\caption{The potential function $\Omega$ for different values of $\mu$}\label{gty}
	\begin{tabular}{|r|c|l|c|c|c|c|}
		\hline
		\textbf{$\ \mu $} & \textbf{$\ \Omega_{C_{1}}$} & \textbf{$\ \Omega_{C_{2}}$} & \textbf{$\  \Omega_{C_{3}}$} \\
		\hline
		\textbf{$\ 0.010 $} & \textbf{$\ 1.56674$} & \textbf{$\  1.48173 $} & \textbf{$\  1.56009$} \\
		\hline
		\textbf{$\ 0.050 $} & \textbf{$\ 1.61362$} & \textbf{$\ 1.40828$} & \textbf{$\  1.58109$}\\
		\hline
		\textbf{$\ 0.100 $} & \textbf{$\ 1.58054$} & \textbf{$\  1.31506 $} & \textbf{$\  1.51818$} \\
		\hline
		\textbf{$\ 0.150 $} & \textbf{$\ 1.49049$} & \textbf{$\  1.21858 $} & \textbf{$\  1.40396$} \\
		\hline
		\textbf{$\ 0.200 $} & \textbf{$\ 1.33994$} & \textbf{$\  1.11573 $} & \textbf{$\  1.24441$} \\
		\hline
		\textbf{$\ 0.240 $} & \textbf{$\ 1.13035$} & \textbf{$\  1.02584 $} & \textbf{$\  1.07758$} \\
		\hline
		\textbf{$\ 0.245 $} & \textbf{$\ 1.08491$} & \textbf{$\  1.01397 $} & \textbf{$\  1.05351$} \\
		\hline
		\textbf{$\ 0.249 $} & \textbf{$\ 1.03081$} & \textbf{$\  1.00435 $} & \textbf{$\  1.03362$} \\
		\hline
	\end{tabular}
\end{table}
\begin{table}
	\centering
	\caption{The Jacobi constant $\acute{C}$ for different values of $\mu$
	}\label{gty}
	\begin{tabular}{|r|c|l|c|c|c|c|}
		\hline
		\textbf{$\ \mu $} & \textbf{$\ \acute{C}=C_{1}$} & \textbf{$\ \acute{C}=C_{2}$} & \textbf{$\  \acute{C}=C_{3}$} \\
		\hline
		\textbf{$\ 0.010 $} & \textbf{$\ 3.13348$} & \textbf{$\  2.96346 $} & \textbf{$\  3.12018$} \\
		\hline
		\textbf{$\ 0.050 $} & \textbf{$\ 3.22724$} & \textbf{$\ 2.81656$} & \textbf{$\  3.16218$}\\
		\hline
		\textbf{$\ 0.100 $} & \textbf{$\ 3.16108$} & \textbf{$\  2.63012 $} & \textbf{$\  3.03636$} \\
		\hline
		\textbf{$\ 0.150 $} & \textbf{$\ 2.98098$} & \textbf{$\  2.43716 $} & \textbf{$\  2.80792$} \\
		\hline
		\textbf{$\ 0.200 $} & \textbf{$\ 2.67988$} & \textbf{$\  2.23146 $} & \textbf{$\  2.48882$} \\
		\hline
		\textbf{$\ 0.240 $} & \textbf{$\ 2.26071$} & \textbf{$\  2.05168 $} & \textbf{$\  2.15516$} \\
		\hline
		\textbf{$\ 0.245 $} & \textbf{$\ 2.16982$} & \textbf{$\  2.02794 $} & \textbf{$\  2.10702$} \\
		\hline
		\textbf{$\ 0.249 $} & \textbf{$\ 2.06162$} & \textbf{$\  2.00870 $} & \textbf{$\  2.06724$} \\
		\hline
	\end{tabular}
\end{table}

\section{N-R BoA, basin entropy and boundary basin entropy}

\subsection{N-R BoA}

We can determine various aspects of  dynamical system with the help of  N-R BoA. Recently, few researchers have applied N-R method in various dynamical system including different perturbing terms in the effective potential (\cite{Dubeibe2020},  \cite{Suraj2019a}, \cite{Suraj2019b}, \cite{Suraj2020} and \cite{Suraj2019},). We have applied N-R iterative scheme (multivariate form) to study the BoA associated with the libration points. To reveal the domain of convergence of a specific libration point, we examine a set of initial conditions. To solve the systems of multivariate function $f(\textbf{x})=0$, we apply the iterative scheme
\begin{align}
{\textbf{x}}_{n+1} ={\textbf{x}}_n - J^{-1}f(\textbf{x}_n),
\end{align} where $f (x_n)$ is the system of equations, while $J^{-1}$ is the corresponding inverse Jacobian matrix.\\
In the present problem, the system of differential equations are given by
\begin{eqnarray*}
\Omega_{x} = 0,\\
\Omega_{y} = 0.
\end{eqnarray*}
With elementary calculations, we get the iterative formula for each coordinate as
\begin{eqnarray}\nonumber
{x}_{n+1}={x}_n-\left(\frac{\Omega_{x_n} \Omega_{{y_n}{y_n}}-\Omega_{y_n} \Omega_{x_ny_n}}{\Omega_{x_nx_n}\Omega_{y_ny_n}-\Omega_{x_ny_n}\Omega_{y_nx_n}}\right), \\
{y}_{n+1}={y}_n+\left(\frac{\Omega_{x_n} \Omega_{{y_nx_n}}-\Omega_{y_n} \Omega_{x_nx_n}}{\Omega_{x_nx_n}\Omega_{y_ny_n}-\Omega_{x_ny_n}\Omega_{y_nx_n}}\right),
\end{eqnarray}
where ${x}_n,{y}_n$ denote the iterates at the $n$-th step of the N-R iterative process. The subscripts denote corresponding partial derivatives of the first and second order of $\Omega(x,y)$.
Partial derivatives of the $\Omega (x,y)$ with respect to $x\ \text{and}\ y$ are as follows
\begin{eqnarray*}
\Omega _x(x, y)=
\end{eqnarray*}
\begin{eqnarray*}
n^2x-\frac{x (1-4 \mu )}{r_0{}^3}-\frac{(-1+x) \mu }{r_1{}^3}-\frac{x \mu }{r_2{}^3}-\frac{(1+x) \mu }{r_3{}^3}-\frac{x
	\mu }{r_4{}^3}
\end{eqnarray*}
\begin{eqnarray*}
\Omega _y(x, y)=
\end{eqnarray*}
\begin{eqnarray*}
n^2y -\frac{y (1-4 \mu )}{r_0{}^3}-\frac{y \mu }{r_1{}^3}-\frac{(y-1) \mu }{r_2{}^3}-\frac{y \mu }{r_3{}^3}-\frac{
	(1+y) \mu }{r_4{}^3}
\end{eqnarray*}
\begin{eqnarray*}
\Omega _{x, x}(x, y)=n^2+\frac{1-4 \mu }{r_0{}^3}\left(\frac{3 x^2 }{r_0{}^2}-1\right)+
\end{eqnarray*}
\begin{eqnarray*}
\frac{ \mu }{r_1{}^3}\left(\frac{3 (x-1)^2 }{r_1{}^2}-1\right)+\frac{
	\mu }{r_2{}^3}\left(\frac{3 x^2 }{r_2{}^2}-1\right)+
\end{eqnarray*}
\begin{eqnarray*}
\frac{ \mu }{r_3{}^3}\left(\frac{3 (x+1)^2 }{r_3{}^2}-1\right)+\frac{ \mu }{r_4{}^3}\left(\frac{3
	x^2 }{r_4{}^2}-1\right)
\end{eqnarray*}
\begin{eqnarray*}
\Omega _{y, y}(x, y)=n^2+\frac{1-4 \mu }{r_0{}^3}\left(\frac{3 y^2 }{r_0{}^2}-1\right)+\frac{ \mu }{r_1{}^3}\left(\frac{3 y^2 }{r_1{}^2}-1\right) +
\end{eqnarray*}
\begin{eqnarray*}
\frac{\mu }{r_2{}^3}\left(\frac{3 (y-1)^2 }{r_2{}^2}-1\right)+\frac{ \mu }{r_3{}^3}\left(\frac{3 y^2 }{r_3{}^2}-1\right)+
\end{eqnarray*}
\begin{eqnarray*}
\frac{ \mu }{r_4{}^3}\left(\frac{3(y+1)^2  }{r_4{}^2}-1\right)
\end{eqnarray*}
\begin{eqnarray}\nonumber
\Omega _{x, y}(x, y)=\frac{3x y (1-4 \mu )}{r_0{}^5}+\frac{3 (x-1) y \mu }{r_1{}^5}+\frac{3 x (y-1) \mu }{r_2{}^5} \\ \nonumber +\frac{3 (1+x) y
	\mu }{r_3{}^5}+\frac{3 x (1+y) \mu }{r_4{}^5}\nonumber
\end{eqnarray}
\begin{eqnarray}\nonumber
\Omega _{y, x}(x, y)=\frac{3x y (1-4 \mu )}{r_0{}^5}+\frac{3 (x-1) y \mu }{r_1{}^5}+\frac{3 x (y-1) \mu }{r_2{}^5} \\ \nonumber +\frac{3 (1+x) y
	\mu }{r_3{}^5}+\frac{3 x (1+y) \mu }{r_4{}^5}\nonumber
\end{eqnarray}
The algorithm of  N-R method is as follows
\begin{itemize}
	\item We choose an initial condition $({x}_0,{y}_0)$ on the configuration plane and apply N-R iterative scheme. In our calculations, we have chosen an uniform grid of $1024\times1024$
	initial conditions (approximately). These initial conditions are called nodes. The minimum and maximum values of ${x}$ and ${y}$ are chosen to view the complete picture of the BoA generated by the libration points.
	\item The method is applied continuously till an accuracy of order $10^{-15}$ ($|x_{n+1}-x_{n}| \leq 10^{-15}$) or the maximum number of iterations (500) is reached.
	\item For each initial conditions, we record the number of iterations $\text{N}$ required to achieve the desired level of accuracy.
	\item We fix the colour for each libration point and initial conditions are assigned with a particular colour according to its convergence towards a specific libration point.
	\item  After assigning all initial conditions with a precise colour, we plot the graph. The graph so obtained is called BoA or basin of convergence. We have used Mathematica (\cite{wolf2017}) to plot the BoA.
\end{itemize}

\subsection{Basin entropy and boundary basin entropy }
In 2016, Daza et. al. (\cite{Alvar16}) had introduced a new tool to measure unpredictability of the BoA. This new tool can quantify the uncertainty of BoA, and it is known as basin entropy. We shall briefly discuss the algorithm for computation of basin entropy:
\begin{itemize}
	\item At first, we complete the process of plotting BoA. We have now each initial condition in the grid of 1024$\times$1024 initial conditions on the configuration plane $(x, y)$ having some colour as per its convergence towards libration points.
	
	\item Now, we divide the whole region into different non-overlapping boxes so that it will completely cover the whole region. Each box should contain precisely 25 trajectories.
	
	\item  We have considered one million trajectories (approximately) for this computation.
	\item We compute the probability of color $j$ inside each box $i$ denoted as $p_{ij}$. The gibbs entropy for every box $i$ is computed as
	\begin{equation}
	\text{S}_{i}= \sum_{j=1}^{m_{i}} \text{p}_{ij} \log\left(\frac{1}{\text{p}_{ij}}\right),
	\end{equation}
	where $m_{i}\in [1,\text{N}_{A}]$, is the number of colors inside the box $i$ and $\text{N}_{A}$ represents the number of libration points (attractors). $\text{p}_{ij}$ is calculated as
	\begin{equation}\nonumber
	\text{p}_{ij}=\frac{\text{number of trajectories leading to color j} }{\text{number of trajectories in the box i}}.
	\end{equation}
	\item We select non overlapping boxes $\text{N}$ so that the total entropy of the grid is equal to the summation of entropy associated to each box $i$.
	\begin{equation*}
	\text{S}=\sum_{i=1}^{\text{N}} \text{S}_{i}= \sum_{i=1}^{\text{N}} \sum_{j=1}^{m_{i}}\text{p}_{ij} \log\left(\frac{1}{\text{p}_{ij}}\right).
	\end{equation*}
	Now, we define the basin entropy as
	\begin{equation*}
	\text{S}_{b}=\frac{\text{S}}{\text{N}}.
	\end{equation*}
	In the same way, we define boundary basin entropy as
	\begin{equation*}
	\text{S}_{bb}=\frac{\text{S}}{\text{N}_{b}}.
	\end{equation*}
where $\text{N}_{b}$ denotes the number of boxes containing more than one color.
\end{itemize}
If the value of $\text{S}_{b}$ and $\text{S}_{bb}$ greater than $\log{2}$ then the BoA or boundaries along BoA is fractal.
Now, we shall study the effect of the parameter $\mu$ on the BoA.

\section{N-R BoA and existence of fractal}

\begin{figure*}[htb!]
	\centering
	\begin{tabular}{cc}
		\includegraphics[scale=.43]{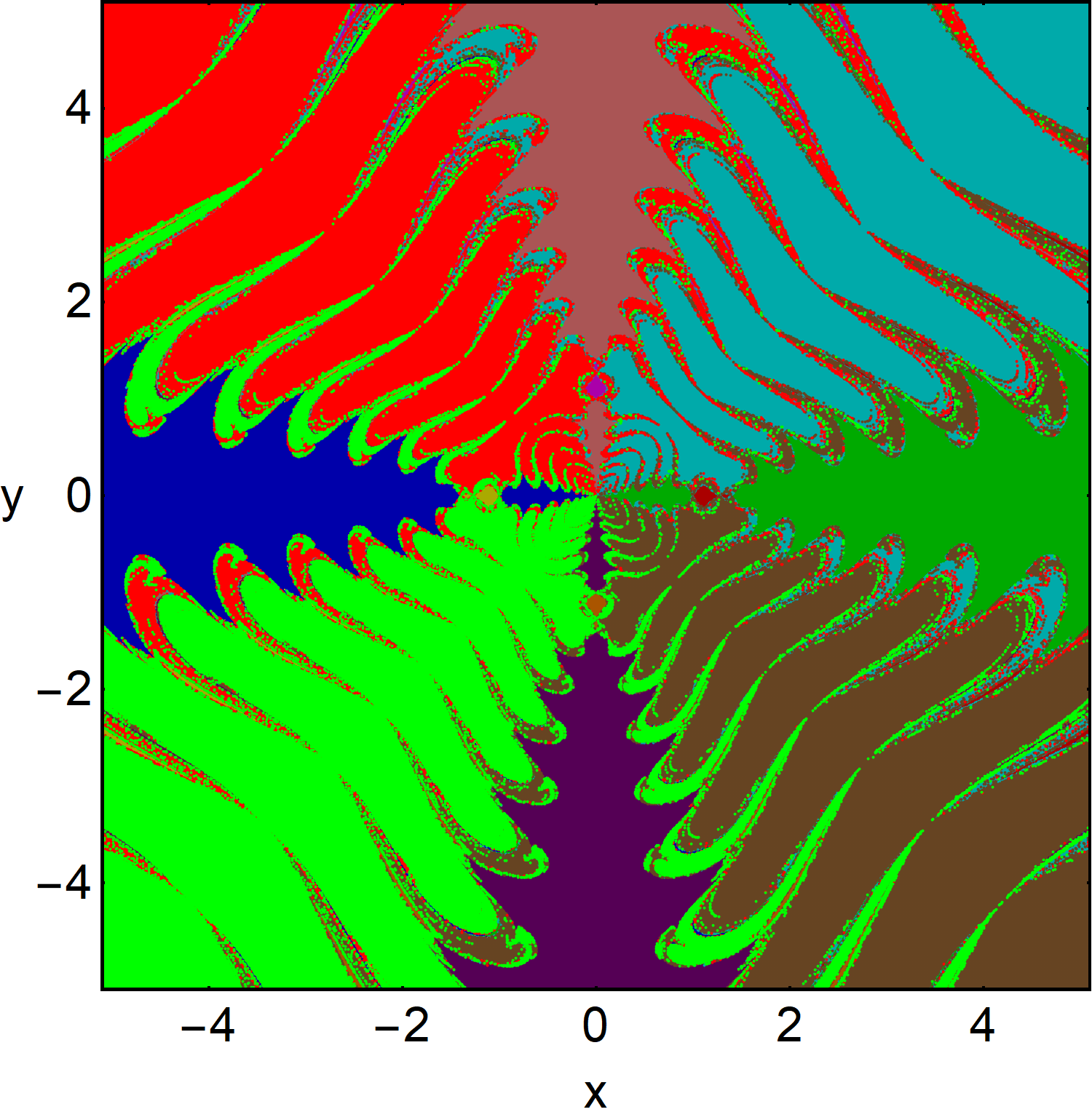}~~~~~~~~~~~~~~&
		\includegraphics[scale=.43]{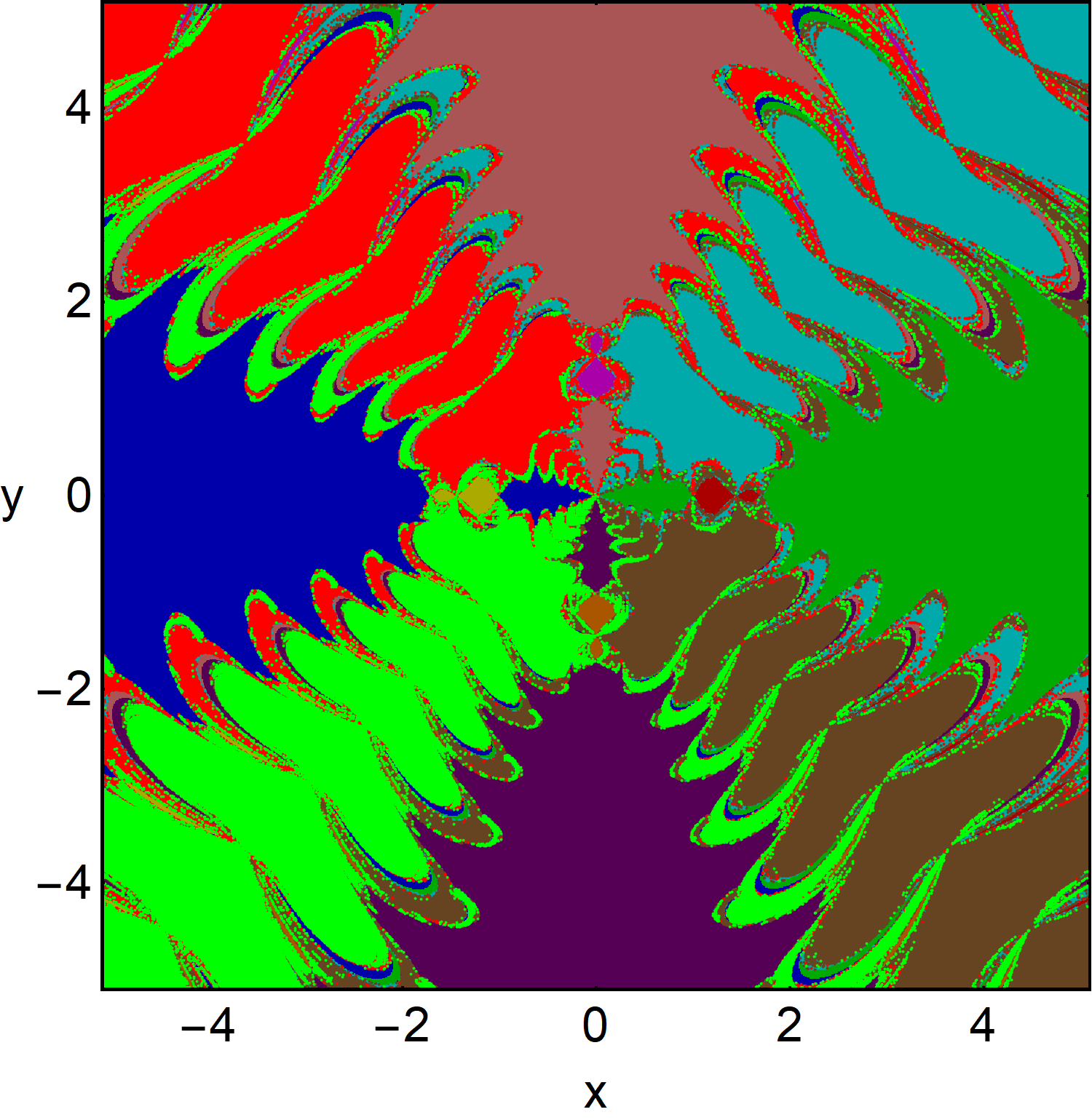}\\ 
		(a) $\mu=$ 0.01~~~~~~&(b) $\mu=$ 0.05\\
		{}&{}\\
		\includegraphics[scale=.43]{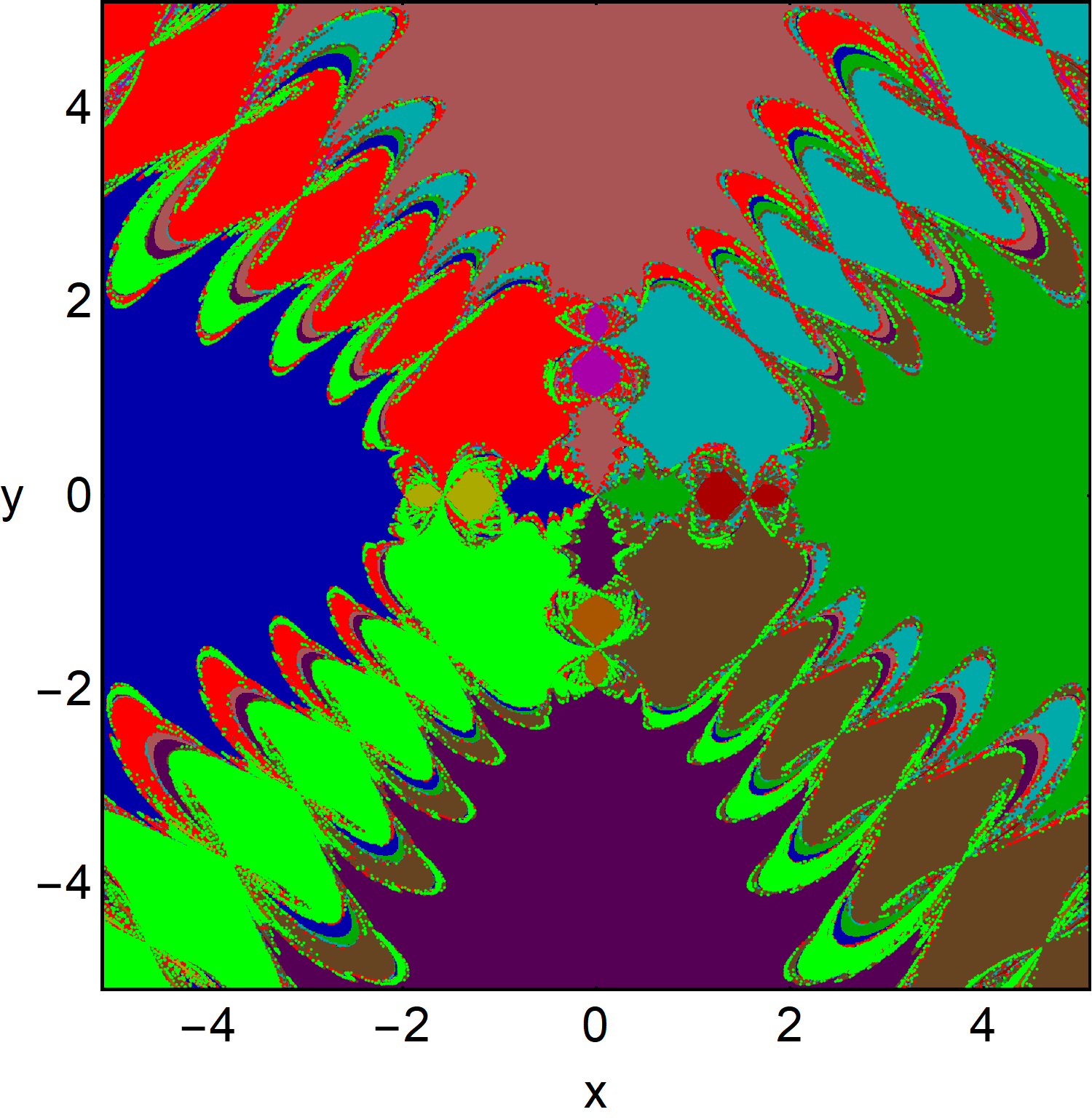}~~~~~~~~~~~~~~&
		\includegraphics[scale=.43]{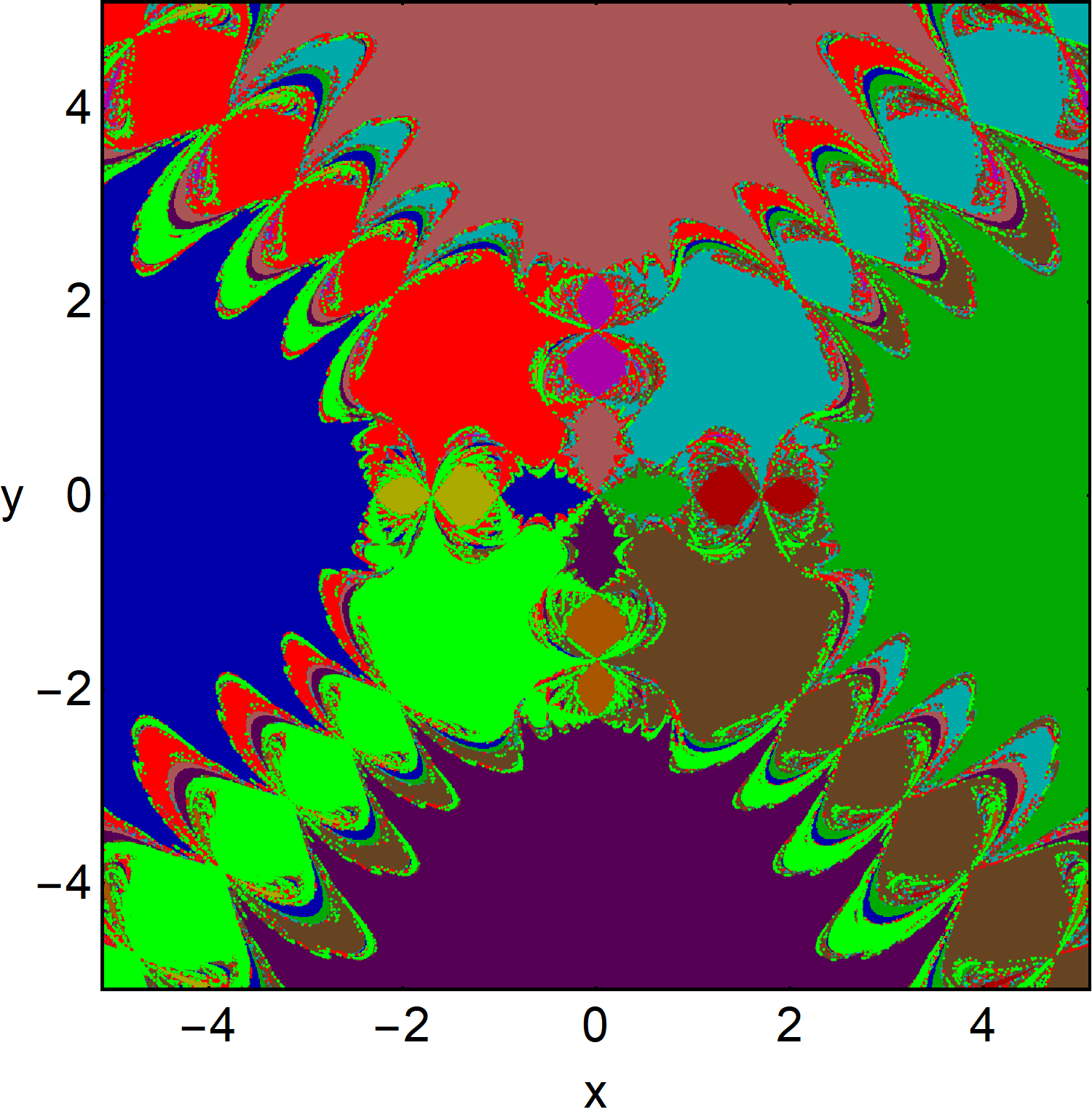}\\ 
		(c)  $\mu=$ 0.1~~~~~~~&(d)  $\mu=$ 0.15\\
		{}&{}\\
	     \includegraphics[scale=.43]{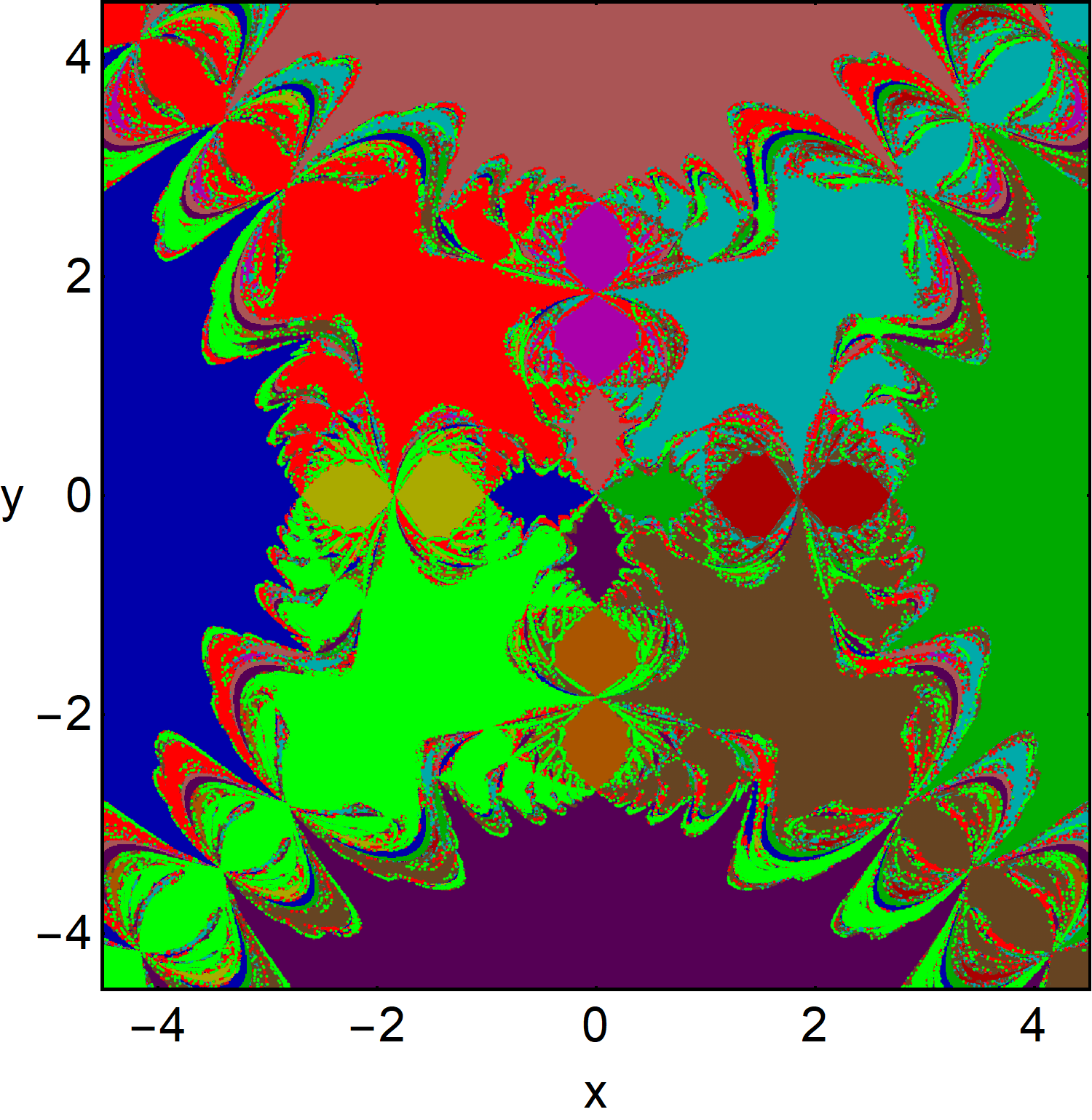}~~~~~~~~~~~~~~&
		\includegraphics[scale=.43]{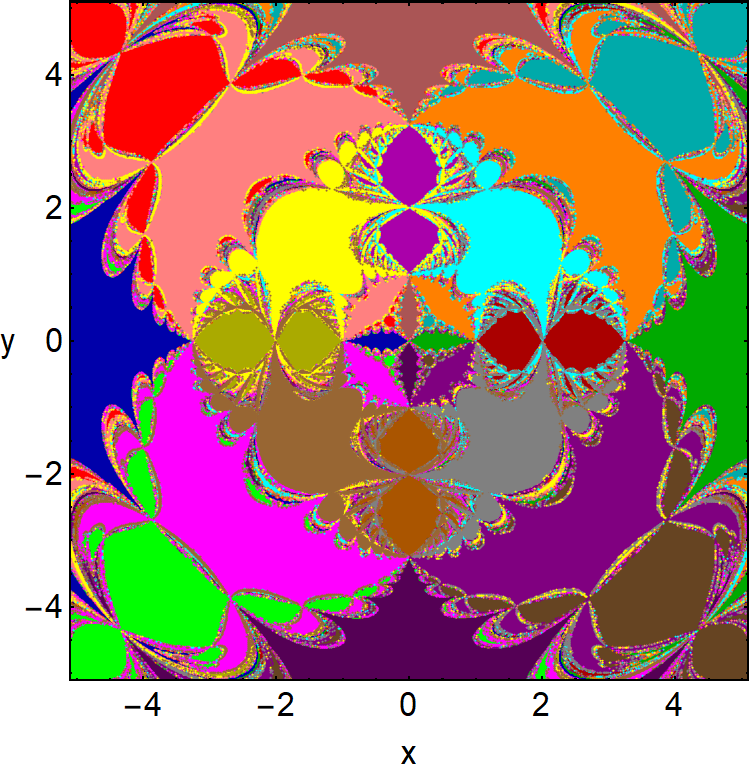}\\
		(e) $\mu=$ 0.2~~~~~~~~ &(f) $\mu=$ 0.24
	\end{tabular}
	\caption{The N-R BoA on the configuration plane $(x, y)$ for different values of the parameter $\mu$ . The colour codes for the BoA corresponding to libration points is as follows: For Fig. (a-e) $L_{1}$ (D. Green), $L_{2}$ (D. Red), $L_{3}$ (D. Blue), $L_{4}$ (D. Yellow), $L_{5}$ (D. Pink), $L_{6}$ (D. Purple), $L_{7}$ (D. Magenta), $L_{8}$ (D. Orange), $L_{9}$ (D. Cyan), $L_{10}$ (D. Brown), $L_{11}$ (Red), $L_{12}$ (Green); (D. stands for darker shade) For Fig. (f) $L_{1}$ (D. Green), $L_{2}$ (D.Red), $L_{3}$ (D.Blue), $L_{4}$ (D.Yellow), $L_{5}$ (D.Pink), $L_{6}$ (D.Purple), $L_{7}$ (D.Magenta), $L_{8}$ (D.Orange), $L_{9}$ (D.Cyan), $L_{10}$ (D.Brown), $L_{11}$ (Red), $L_{12}$ (Green), $L_{13}$ (Orange), $L_{14}$ (Purple), $L_{15}$ (Pink), $L_{16}$ (Magenta), $L_{17}$ (Cyan), $L_{18}$ (Gray), $L_{19}$ (Yellow), $L_{20}$ (Brown)}
\end{figure*}
\begin{figure*}[htb!]
	\centering
	\begin{tabular}{cc}
		\includegraphics[scale=.50]{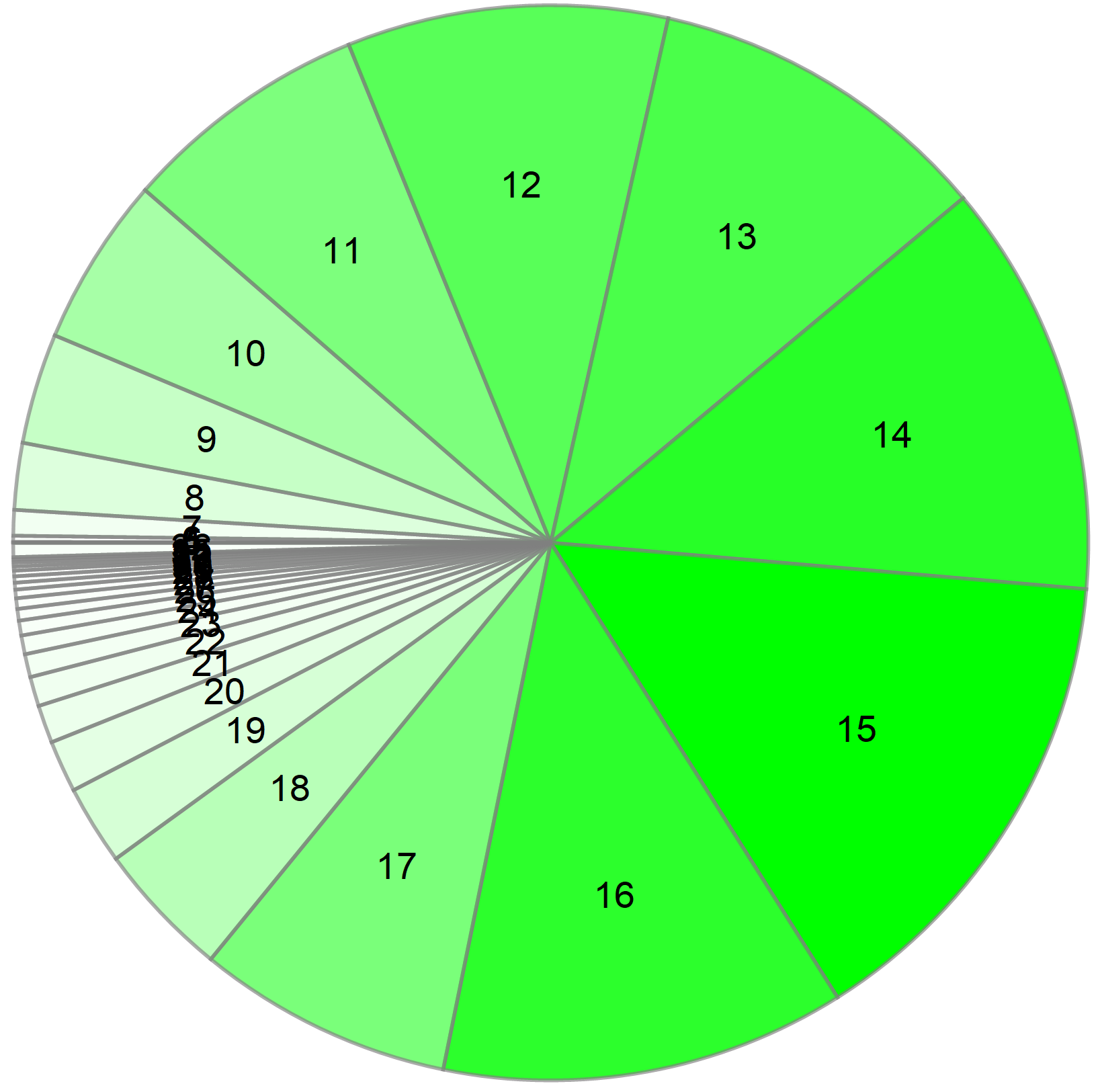}~~~~~~~~~~~~&
		\includegraphics[scale=.50]{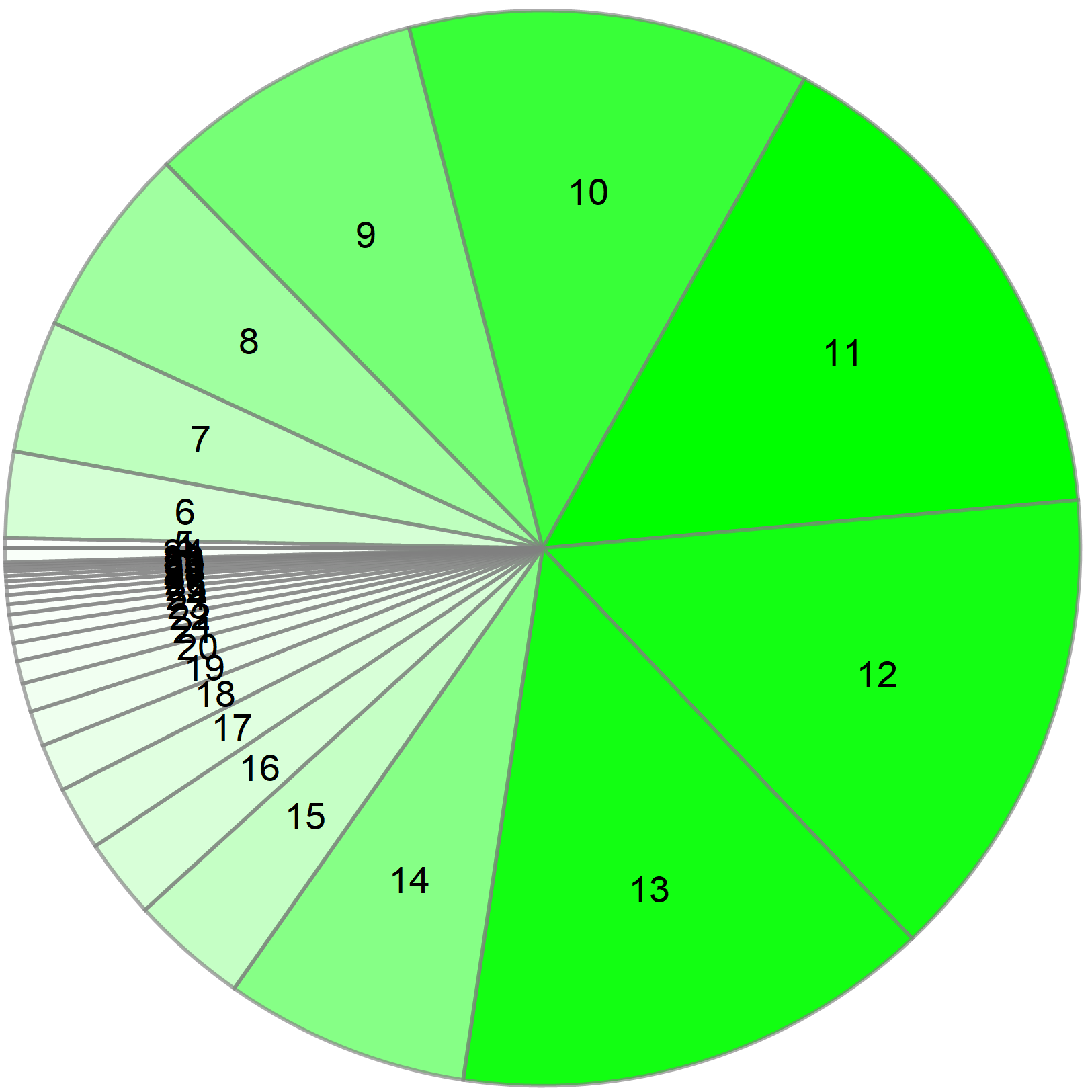}\\
		(a) $\mu=$ 0.01~~~~~~~~&(b) $\mu=$ 0.05\\
			{}&{}\\
		\includegraphics[scale=.50]{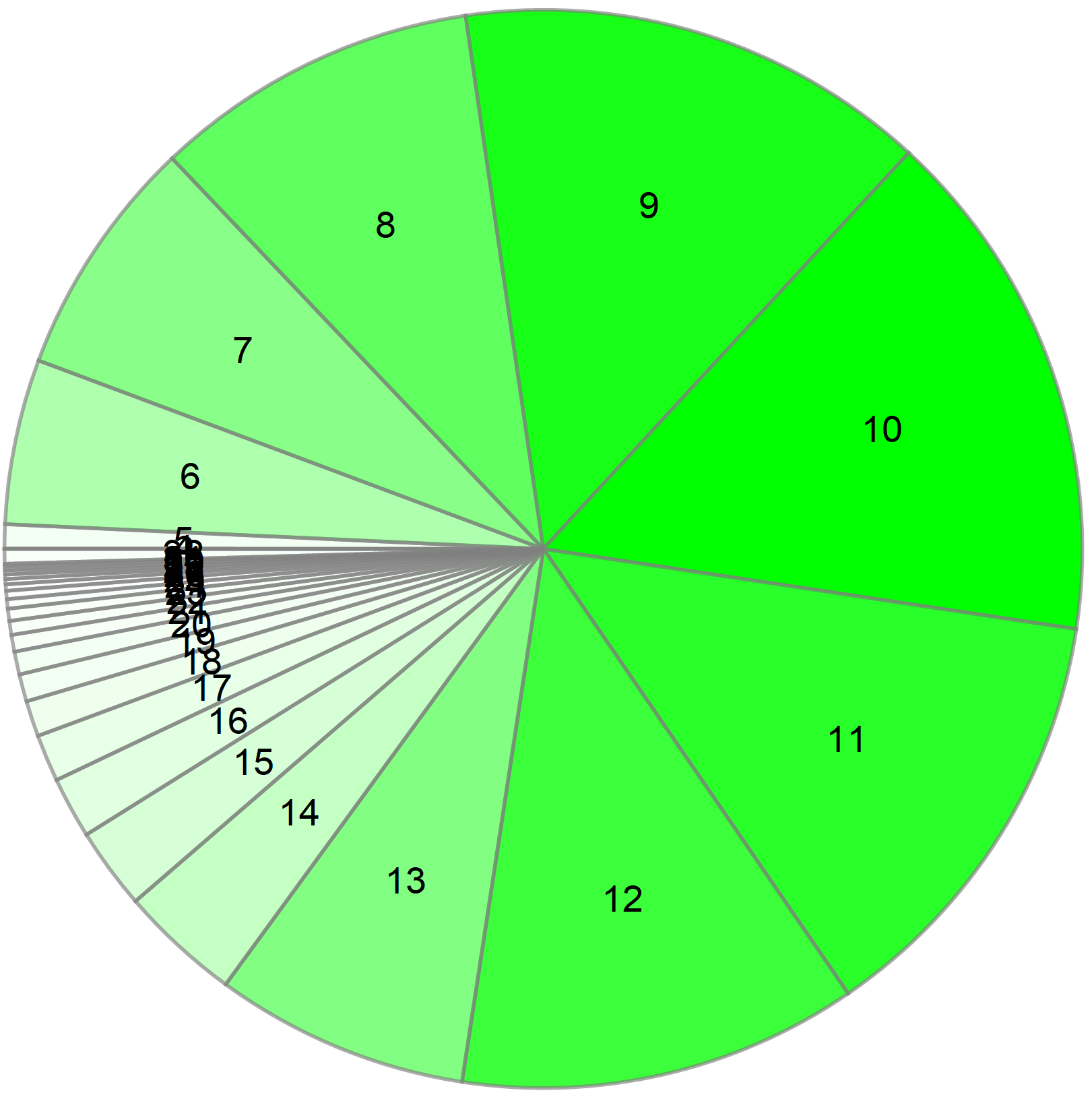}~~~~~~~~~~~~~&
		\includegraphics[scale=.50]{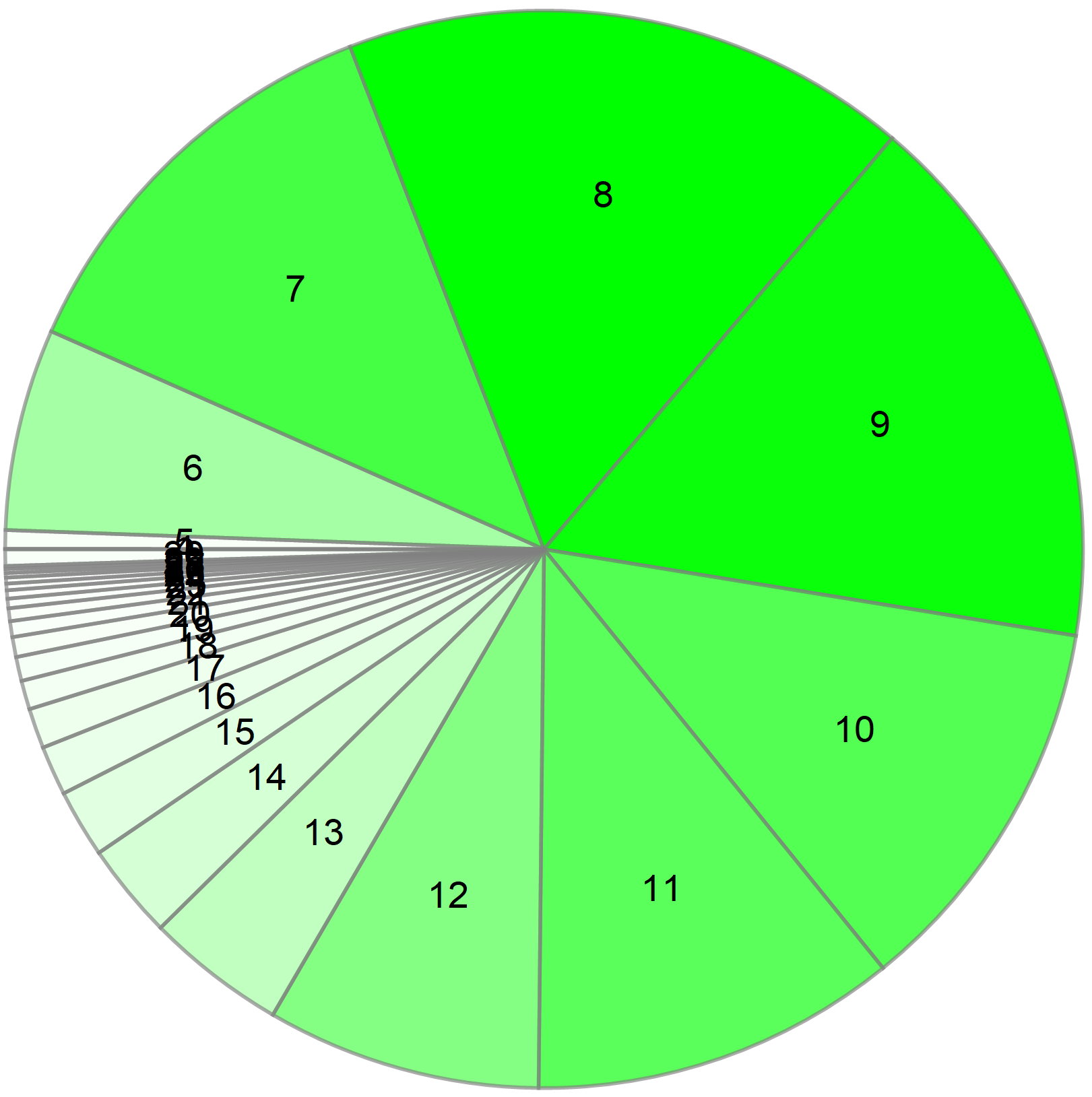}\\
		(c)  $\mu=$ 0.1~~~~~~~~&(d)  $\mu=$ 0.15\\
		{}&{}\\
		\includegraphics[scale=.50]{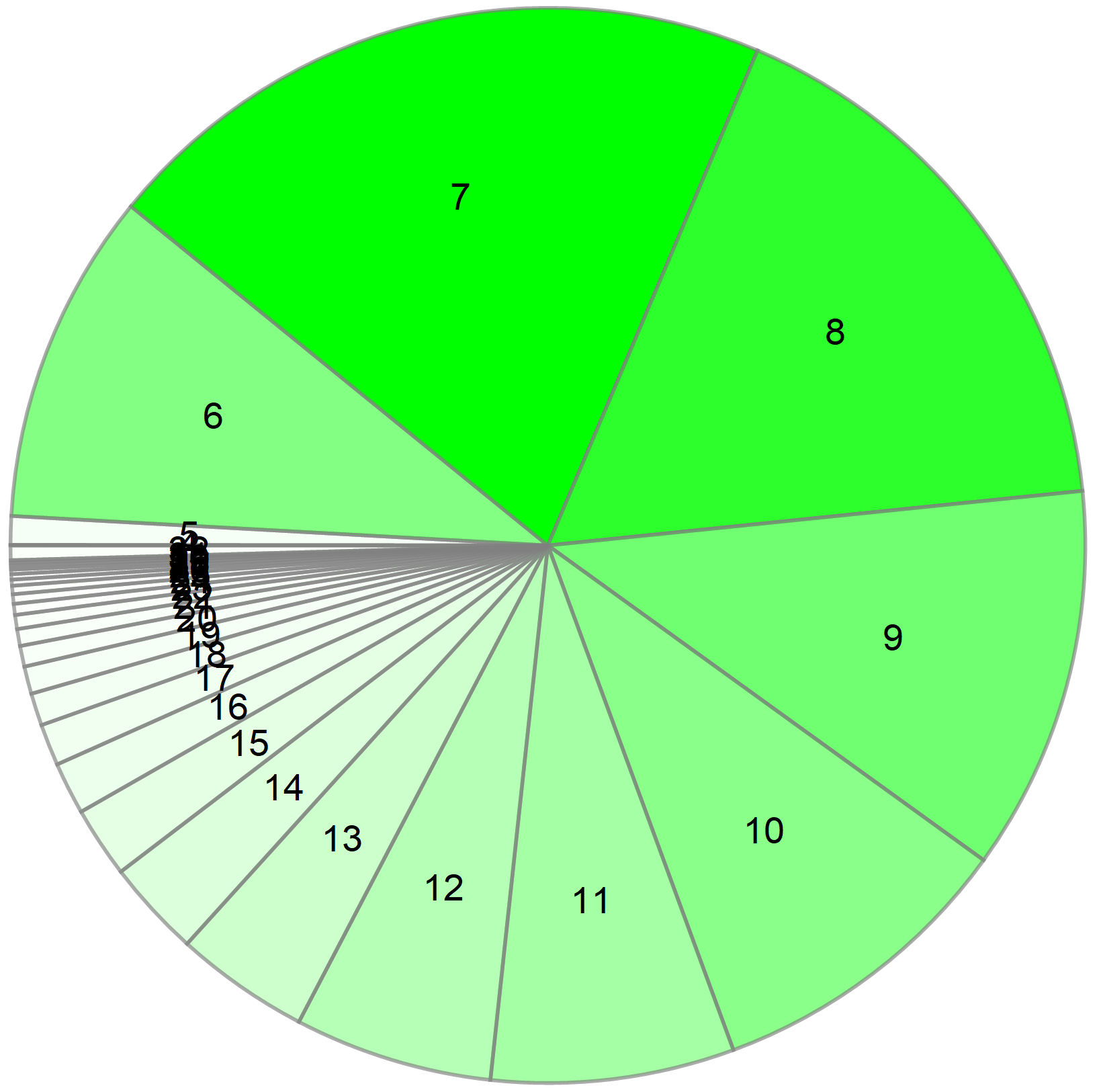}~~~~~~~~~~~~~&
		\includegraphics[scale=.50]{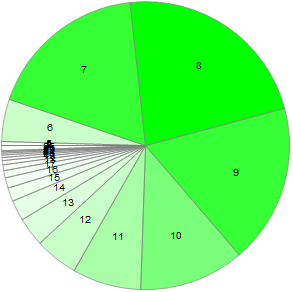}\\
		(e) $\mu=$ 0.2 ~~~~~~~~&(f)$\mu=$ 0.24
	\end{tabular}
	\caption{Pi-Chart representing the number of iterations needed for the convergence of initial conditions toward libration points. Each sector with given number 'N' denote the proportion of initial conditions converges after 'N' iterations.}
\end{figure*}
\begin{figure*}[htb!]
	\centering
	\begin{tabular}{cc}
		\includegraphics[scale=.43]{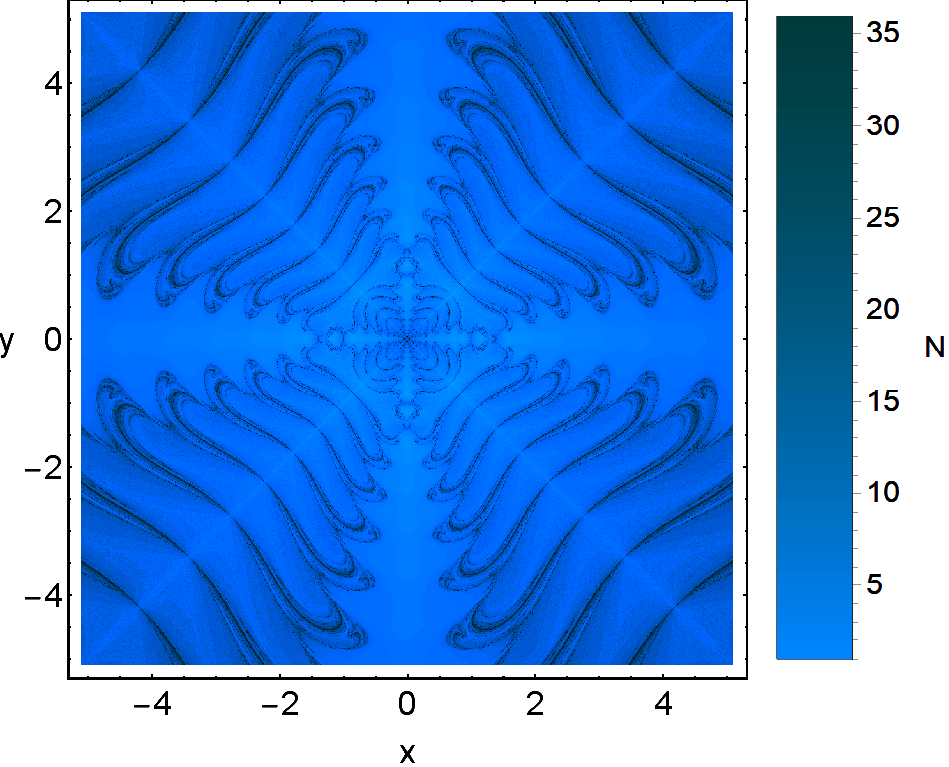}~~~~~~~~&
		\includegraphics[scale=.43]{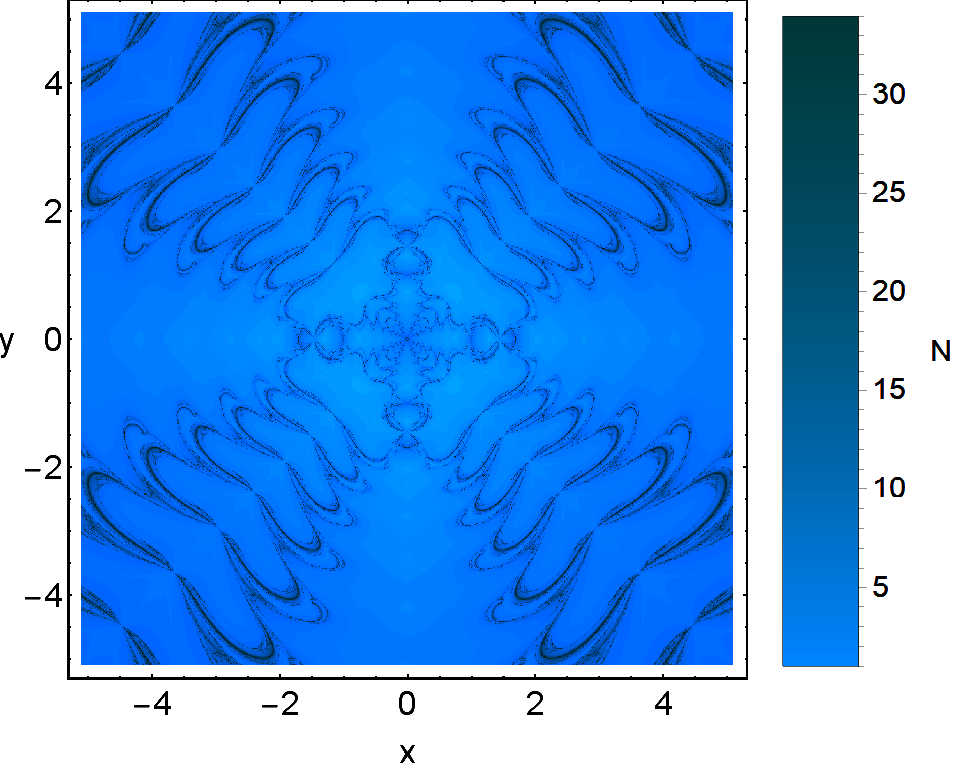}\\
		(a) $\mu=$ 0.01~~~~~~~~~~~~~~~~&(b) $\mu=$ 0.05~~~~\\
		{}&{}\\
		\includegraphics[scale=.43]{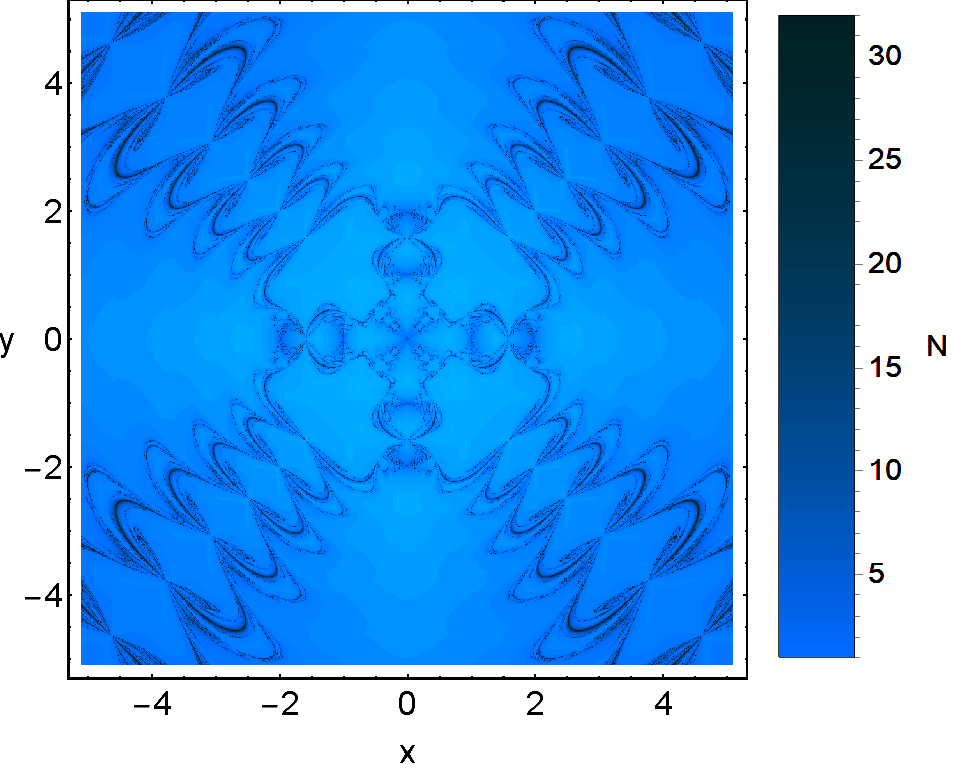}~~~~~~~~&
		\includegraphics[scale=.43]{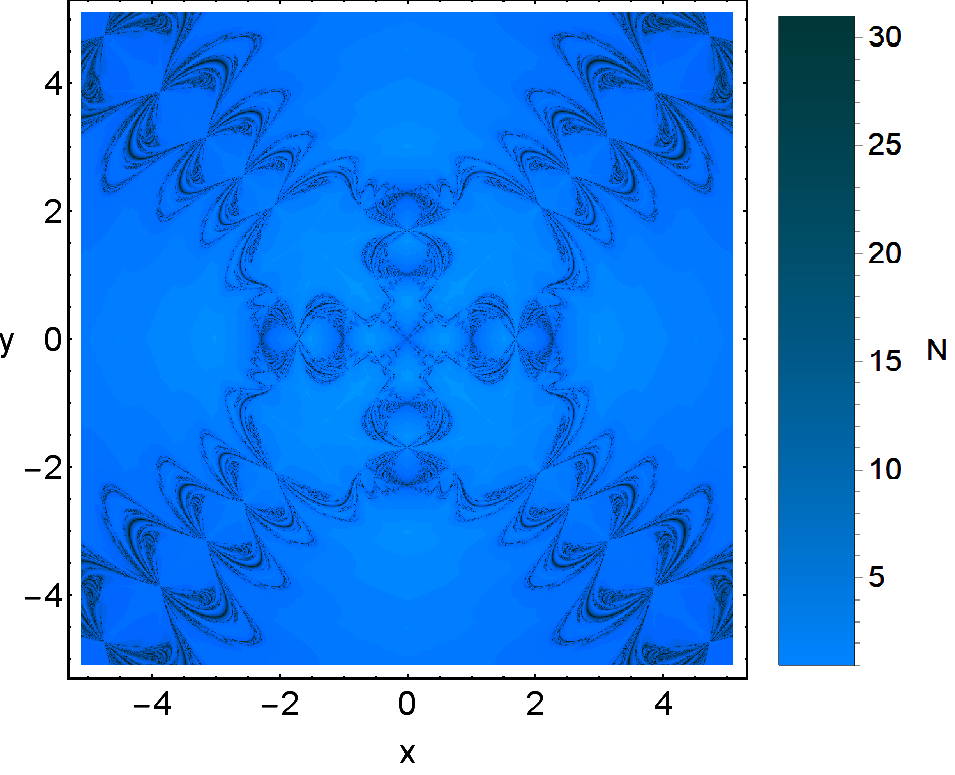}\\
		(c)  $\mu=$ 0.1~~~~~~~~~~~~~~~~~&(d)  $\mu=$ 0.15~~~~\\
		{}&{}\\
		\includegraphics[scale=.43]{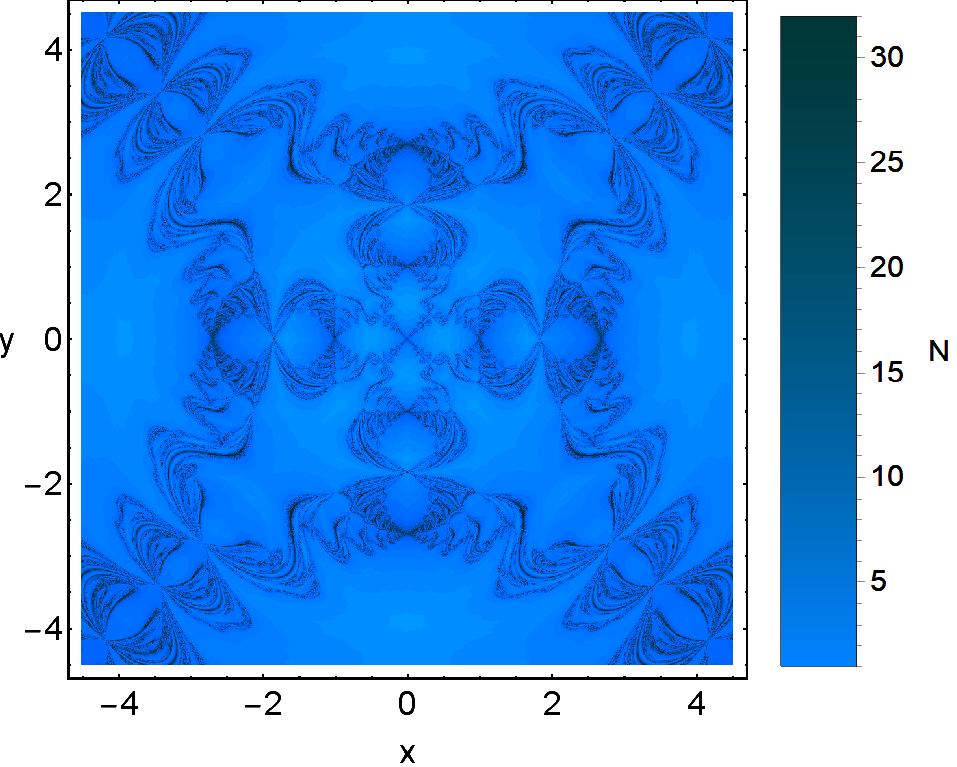}~~~~~~~~&
		\includegraphics[scale=.43]{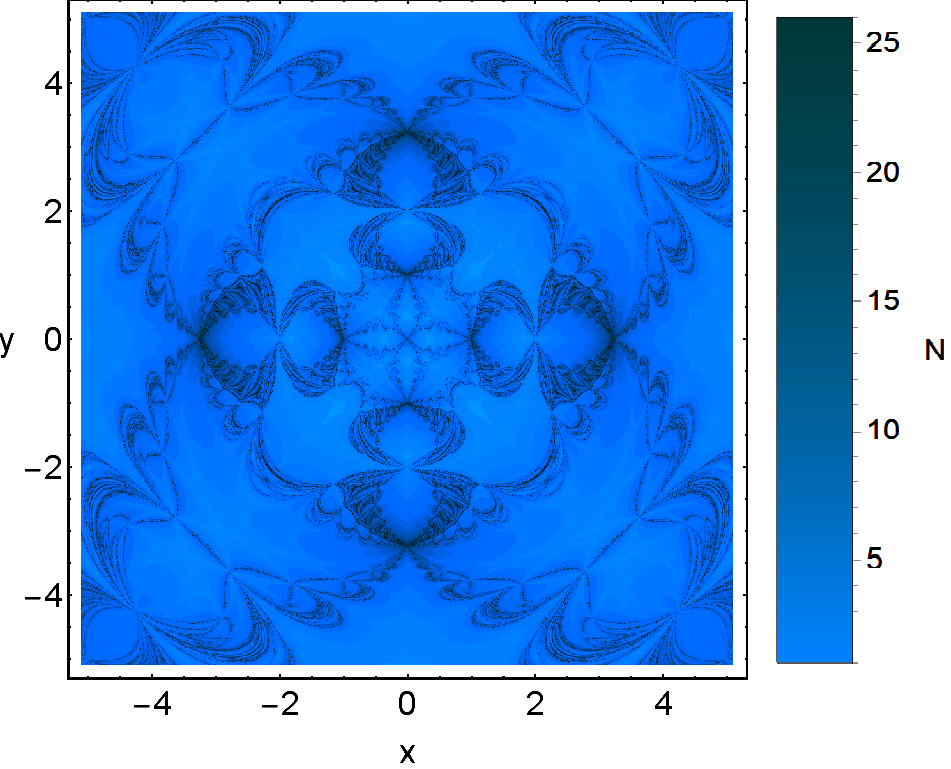}\\
		(e) $\mu=$ 0.2 ~~~~~~~~~~~~~~~~~&(f) $\mu=$ 0.24~~~~~
	\end{tabular}
	\caption{Number of N-R iterations required to achieve the desired accuracy is shown using the intensity of blue colour tone. Initial conditions located at boundaries needed comparatively more number N-R iterations.}
\end{figure*}

\begin{table}
	\centering
	\caption{Distribution of initial conditions toward libration points $L_{i}, i=1,2,3,...,12$ for $\mu$=0.01, 0.05, 0.10, 0.15, 0.20.}\label{gty}
	\begin{tabular}{|r|c|l|c|c|c|c|}
		\hline
		\textbf{$\ \mu $} & \textbf{$\ 0.01$} & \textbf{$\ 0.05$} & \textbf{$\  0.10$} & \textbf{$\  0.15$}& \textbf{$\  0.20$}\\
		\hline
		\textbf{$\ L_{1} $} & \textbf{$\ 64013$} & \textbf{$\  110983 $} & \textbf{$\  137228$} & \textbf{$\  150423$}& \textbf{$\  148072$}\\
		\hline
		\textbf{$\ L_{2} $} & \textbf{$\ 6340$} & \textbf{$\ 7639$} & \textbf{$\  8898$} & \textbf{$\  12856$}& \textbf{$\  19594$}\\
		\hline
		\textbf{$\ L_{3} $} & \textbf{$\ 64014$} & \textbf{$\  110976 $} & \textbf{$\  137223$} & \textbf{$\  150424$}& \textbf{$\  148059$}\\
		\hline
		\textbf{$\ L_{4} $} & \textbf{$\ 6340$} & \textbf{$\  7641 $} & \textbf{$\  7641$} & \textbf{$\  12854$}& \textbf{$\  19589$}\\
		\hline
		\textbf{$\ L_{5} $} & \textbf{$\ 64013$} & \textbf{$\  110980 $} & \textbf{$\  137229$} & \textbf{$\  150424$}& \textbf{$\  148068$}\\
		\hline
		\textbf{$\ L_{6} $} & \textbf{$\ 64012$} & \textbf{$\  110980 $} & \textbf{$\  137222$} & \textbf{$\  150431$}& \textbf{$\  148057$}\\
		\hline
		\textbf{$\ L_{7} $} & \textbf{$\ 6341$} & \textbf{$\  7641 $} & \textbf{$\  8898$} & \textbf{$\  12854$}& \textbf{$\  19593$}\\
		\hline
		\textbf{$\ L_{8} $} & \textbf{$\ 6342$} & \textbf{$\  7641 $} & \textbf{$\  8903$} & \textbf{$\  12853$}& \textbf{$\  19588$}\\
		\hline
		\textbf{$\ L_{9} $} & \textbf{$\ 190256$} & \textbf{$\  141994 $} & \textbf{$\  114489$} & \textbf{$\  97334$}& \textbf{$\  92954$}\\
		\hline
		\textbf{$\ L_{10} $} & \textbf{$\ 190261$} & \textbf{$\  141993 $} & \textbf{$\  114487$} & \textbf{$\  97336$}& \textbf{$\  92963$}\\
		\hline
		\textbf{$\ L_{11} $} & \textbf{$\ 190261$} & \textbf{$\ 141994 $} & \textbf{$\  114489$} & \textbf{$\  97336$}& \textbf{$\  92961$}\\
		\hline
		\textbf{$\ L_{12} $} & \textbf{$\ 190260$} & \textbf{$\  141991 $} & \textbf{$\  114485$} & \textbf{$\  97328$}& \textbf{$\  92954$}\\
		\hline
	\end{tabular}
\end{table}
\begin{table}
	\centering
	\caption{Distribution of initial conditions toward libration points $L_{i}, i=1,2,3,...,20$ for $\mu$=0.24.}\label{gty}
	\begin{tabular}{|r|c|l|c|c|c|c|}
		\hline
		\textbf{$\ \mu $} & \textbf{$\ 0.24$} & {$\ \mu $} & \textbf{$\ 0.24$} & {$\ \mu $} & \textbf{$\ 0.24$}\\
		\hline
		\textbf{$\ L_{1} $} & \textbf{$\ 64637$} & {$\ L_{8} $} & \textbf{$\ 25195$} & {$\ L_{15} $} & \textbf{$\ 82702$}\\
		\hline
		\textbf{$\ L_{2} $} & \textbf{$\ 25192$} & {$\ L_{9} $} & \textbf{$\ 49930$} & {$\ L_{16} $} & \textbf{$\ 82704$}\\
		\hline
		\textbf{$\ L_{3} $} & \textbf{$\ 64644$} & {$\ L_{10} $} & \textbf{$\ 49924$} & {$\ L_{17} $} & \textbf{$\ 38148$}\\
		\hline
		\textbf{$\ L_{4} $} & \textbf{$\ 25197$} & {$\ L_{11} $} & \textbf{$\ 49925$} & {$\ L_{18} $} & \textbf{$\ 38150$}\\
		\hline
		\textbf{$\ L_{5} $} & \textbf{$\ 64638$} & {$\ L_{12} $} & \textbf{$\ 49927$} & {$\ L_{19} $} & \textbf{$\ 38150$}\\
		\hline
		\textbf{$\ L_{6} $} & \textbf{$\ 64645$} & {$\ L_{13} $} & \textbf{$\ 82706$} & {$\ L_{20} $} & \textbf{$\ 38151$}\\
		\hline
		\textbf{$\ L_{7} $} & \textbf{$\ 25194$} & {$\ L_{14} $} & \textbf{$\ 82702$} & {} & \textbf{}\\
		\hline
	\end{tabular}
\end{table}
In this section, we will discuss the BoA for different values of the parameter $\mu$. We have chosen $\mu$ in the interval (0, 0.25). Results regarding all the cases are presented in Table 3 and Figs. (4, 5, 6 (a-f)). We will discuss each case to understand the impact of a parameter on BoA.

\subsection{Case of twelve libration points}
The values of parameter $\mu$, in this case are 0.01, 0.05, 0.1, 0.15 and 0.2 respectively (Table 3). The domain of convergence of each libration points is infinite. BoA is in well-defined shape for each case.  There are twelve ($L_{i}, i=1, 2, 3...12$) libration points in this case. In Figs. 4(a-e), the BoA obtained using multivariate N-R method is shown. We consider  approx one million initial conditions on configuration plane $(x,y)$. We notice that all initial conditions converge to some libration points. We do not observe any non-converging initial condition in any case. The data regarding convergence of all initial conditions towards libration points is shown in the Table 3. The well formed BoA covers all of the configuration planes. The domain of convergence for all ($L_{i},  i=1, 2, 3, ..12.$) extend towards infinity. Basins are in a symmetrical shape concerning $x$-axis and $y$-axis. The domain of convergence is intertwined along the boundaries. As we increase the value of $\mu$, we see that the central region of BoA is getting zoomed-in.  We notice a remarkable change in the number of initial conditions converging towards libration points due to change in the parameter. In Fig. (4a) Approx 1,90,260 initial conditions converges towards $L_{9}, L_{10}, L_{11}$ and  $L_{12}$. Approximately sixty five thousands initial conditions converge towards $L_{1}, L_{3}, L_{5}$ and  $L_{6}$. Nearly six thousand initial condition converge towards $L_{2}, L_{4}, L_{7}$ and  $L_{8}$. Thus, for this case  $L_{9}, L_{10}, L_{11}$ and $ L_{12}$ are said to be strong libration points (attractors). On the other hand, $L_{2}, L_{4}, L_{7}$ and $ L_{8}$ are said to be weak libration points (attractors).

Now, for the next case shown in Fig. (4b), the value of the parameter $\mu$ is slightly increased.  The shape in the form of four strips going away from the origin is comparatively less wider in this case. Due to variation in $\mu$, there is a notable change in the number of initial conditions converging towards libration points. In this case, similar to the previous case,  $L_{2}, L_{4}, L_{7}$ and $ L_{8}$ are the weak attractors attracting approximately seven thousand five hundred initial conditions. Initial conditions (approx one lakhs forty two thousands) converge towards  $L_{9}, L_{10}, L_{11}$ and $ L_{12}$ (strong attractors in this case). Now we give a little increase in the parameter $\mu$ and it is 0.1 now. Contrary to the previous case, $L_{1}, L_{3}, L_{5}$ and $ L_{6}$ are the strong attractors attracting one lakh thirty-seven thousand initial conditions. In this case, $L_{2}, L_{4}, L_{7}$ and $ L_{8}$ attracts eighty-nine thousand are weak libration points . Similar to these two cases in other three cases, we notice a remarkable change in the number of initial conditions converge towards libration points.

Now we shall discuss the data related to BoA presented in Figs. 5 (a-e) in the form of Pi-Charts. We have shown the number of iterations of N-R method needed for the convergence of initial conditions (Approx. one million). In Fig. (5a), we see that 95\% initial conditions converge after 36 iterations. The maximum number of initial conditions (exactly 152904 initial conditions) converge after fifteen (15) iterations. Also, it is essential to note that one in one million initial conditions converge after 80th iteration (maximum iteration). In Fig. (5b), we see that 95\% initial conditions converge after 34 iterations. The maximum number of initial conditions (exactly 161318 initial conditions) converge after eleven iteration. Also, one in one million initial conditions converges after the 88th iteration (maximum iteration). In Fig. (5c), we see that 95\% initial conditions converge after 32 iterations. The maximum number of initial conditions (exactly 161719 initial conditions) converge after twenty one  iteration. We notice that one particular initial condition in one million initial conditions converge after 80th iteration (maximum iteration needed in this case).

In Fig. (5d), we see that 95\% initial conditions converge after 30 iterations. The maximum number of initial conditions (exactly 1,77,642 initial conditions) converge after twentieth (8) iterations. Also, it is essential to note that one in one million initial conditions converge after 80th iteration (maximum iteration). In Fig. (5e), we see that 95\% initial conditions converge after 39 iterations. The maximum number of initial conditions (exactly 2,13,817 initial conditions) converge after eighteenth  (7th) iteration. Also, one in one million initial conditions converges after eighty-two (86th) iteration (maximum iteration).

Now, we discuss the cases given in Figs. 6 (a-e). These figures are meaningful in the sense that they locate the initial conditions in the shades of blue colour as per the number of iterations of the N-R method needed to achieve the desired accuracy. In Figs. 6 (a-e), we can see that the initial conditions lying away from the boundary lines of different BoA need less number of iterations as compared to the initial conditions lying along the boundaries. Dark shed of blue colour corresponds to initial conditions which require a higher number of iterations for their convergence. Thus merely the visualization of these graphs gives us the idea that it is a complex behaviour.

\subsection{Case of twenty libration points}
 The value of parameter $\mu$, in this case, is 0.24. There are twenty libration points ($L_{i}$, i=1, 2, 3,...,20) in this case (Fig. 2b). In Fig. 4 (f), we have shown the BoA. We have considered one million initial conditions approximately in the grid of 1024 \time 1024 in the configuration plane $(x, y)$. Similar to previous cases, we do not encounter any non-converging initial conditions. The data regarding convergence of initial conditions towards libration points is shown in Table 4. The domain of convergence of all libration points (attractors) extends to infinity. The BoA covers the whole configuration plane and well defined in shape. In this case, $L_{13}, L_{14}$, $L_{15}$ and $L_{15}$ are the strong attractors attracting eighty two thousand initial conditions approx. $L_{2}, L_{4}, L_{7}$ and $L_{8}$ are the weak attractors attract twenty five thousand initial conditions approx. In Fig. 5(f), Pi-Chart related to BoA is presented. We find that 95\% initial conditions converge after 25 iterations of the N-R method. The maximum number of initial conditions (exactly 2,35,048 initial conditions) converge after twentieth (8th) iteration. Also, it is important to note that two initial conditions in one million initial conditions converge after eighty fourth iteration (maximum iteration).

In Fig. 6 (f), we notice that initial conditions lying along the boundary line of BoA need iterations more than the initial conditions lying away from the boundary. This result is similar to previous cases. Now we will discuss the degree of unpredictability in the BoA.

\subsection{Basin entropy and boundary basin entropy}

\begin{figure*}[htb!]
	\begin{tabular}{cc}
		 \includegraphics[scale=.63]{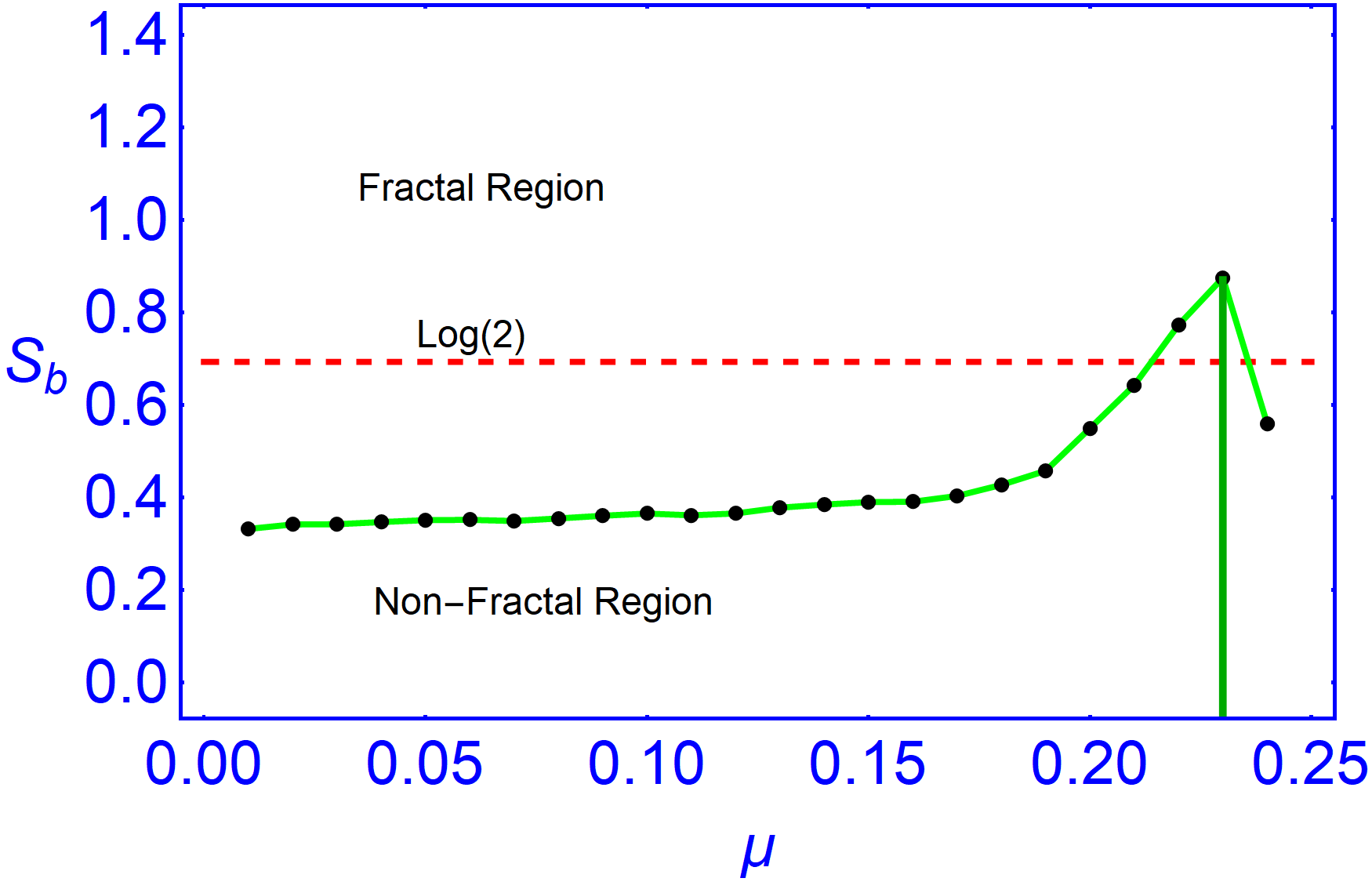}~~~~~~~~&\includegraphics[scale=.63]{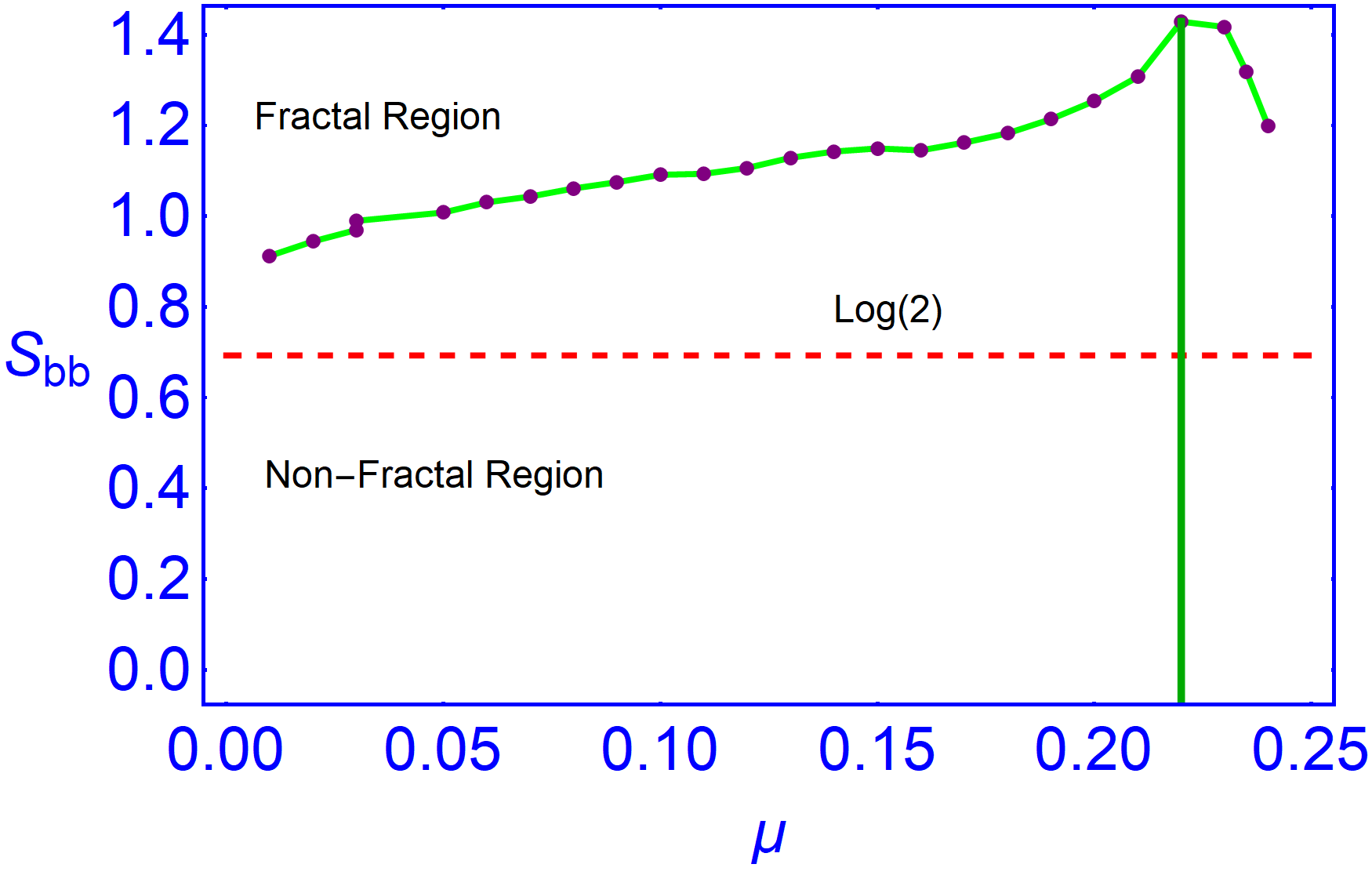}\\
		(a)  $\mu$ versus $S_{b}$~~& (b) $\mu$ versus $S_{bb}$
	\end{tabular}
	\caption{(a) Graphs of the basin entropy $S_{b}$ for $\mu \in (0, 0.25)$ (b) Graphs of the basin entropy $S_{bb}$ for $\mu \in (0, 0.25)$. In both figures, the dotted line divides the whole region into two parts; fractal region and non-fractal region.}
\end{figure*}

\begin{figure*}[htb!]
	\begin{tabular}{cc}
		\includegraphics[scale=.63]{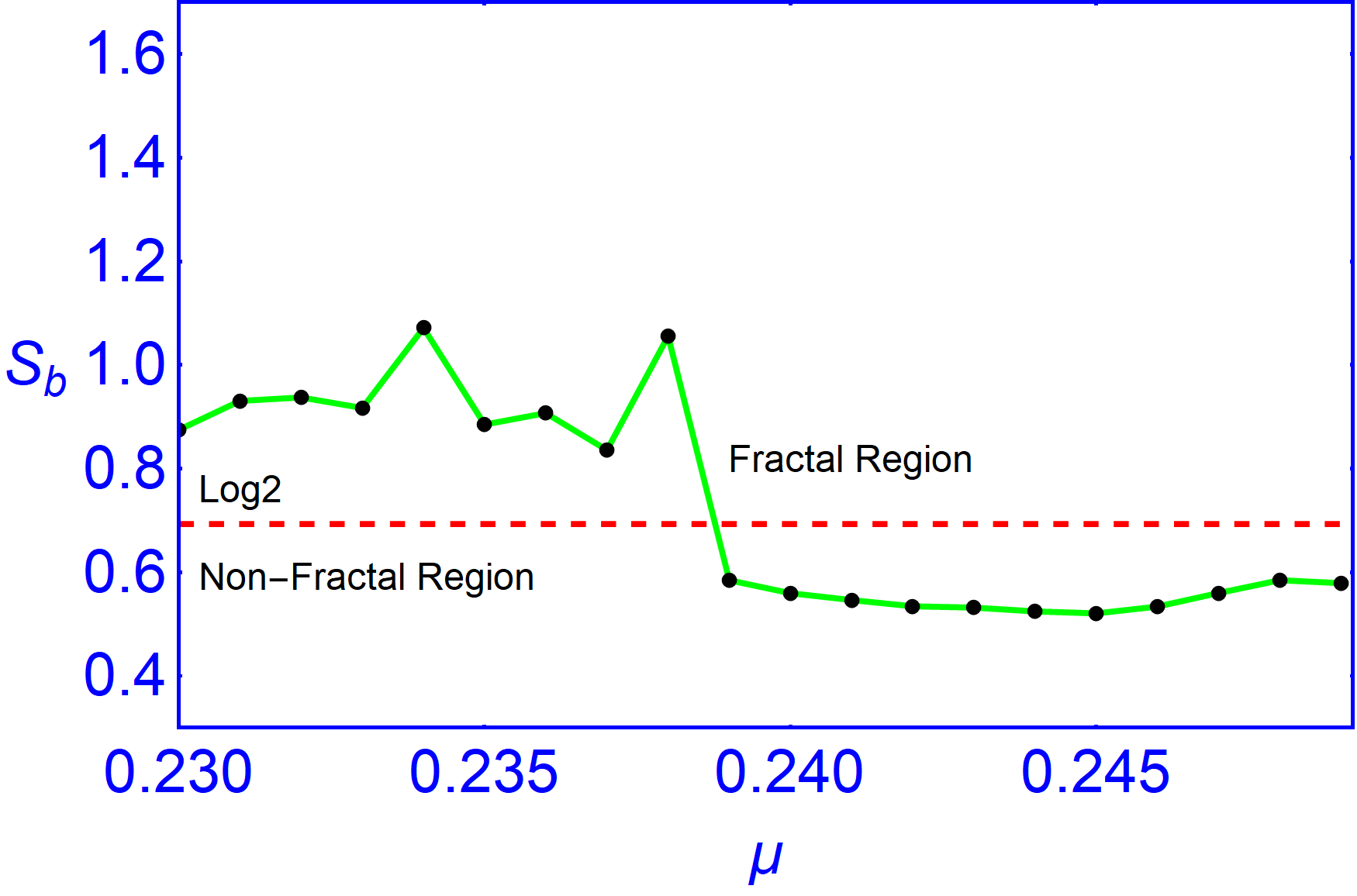}~~~~~~~~&\includegraphics[scale=.63]{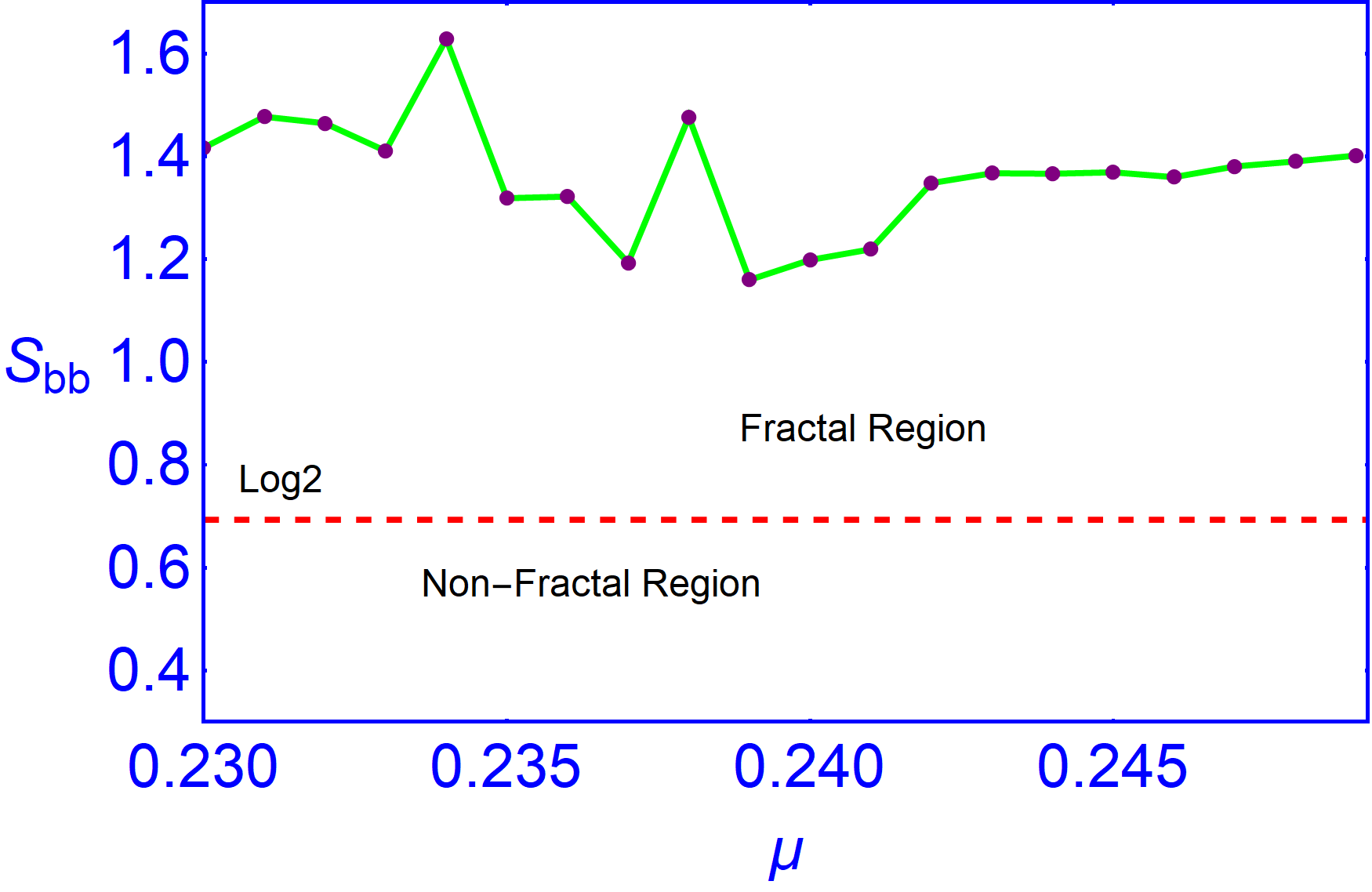}\\
		(a)  $\mu$ versus $S_{b}$~~& (b) $\mu$ versus $S_{bb}$
	\end{tabular}
	\caption{(a) Graph of the basin entropy $S_{b}$ for $\mu \in (0.23, 0.25)$  (zoom-in of Fig. 7(a))  (b) Graph of the boundary basin entropy $S_{bb}$ for $\mu \in (0.23, 0.25)$  (zoom-in of Fig. 7(b)). In both figures, the dotted line divides the whole region into two parts; fractal region and non-fractal region.}
\end{figure*}
  We have investigated the unpredictable behaviour of the BoA or the existence of a fractal region in BoA or along the boundaries using the method of basin entropy. For several values of the parameter mass ratio ($\mu$), the values of basin entropy and boundary basin entropy are calculated using the algorithm mentioned in Section 3. We have plotted two graphs in Figs. 7. Fig. 7 (a) gives the idea of unpredictability in the BoA as a whole. Fig. 7 (b) tells us about the existence of fractal region along the boundaries of BoA. Here, the existence of fractal or the unpredictability of BoA has a particular meaning. We observe that along the boundaries of BoA or sometimes throughout the basin, the domain of convergence of libration points are highly intermixed. In that case, the final destination of the initial condition is highly sensitive. It simply means that the two very close initial conditions may converge to two different libration points. Thus, the sensitivity of initial conditions of BoA is also a matter of investigation.

  In Fig. 7(a), we observe that for the values of $\mu$ from 0.01 to 0.21 the value of $S_{b}$ lies below the fractal region. For the value of $\mu$=0.22 and 0.23, the value of $S_{b}$ lies in the fractal region. For the value of $\mu$=0.24, again the basin entropy lies in the non-fractal region. Thus, for only the values of $\mu$ between 0.22 to 0.23, the existence of fractal is confirmed for the BoA. In Fig. 7(b), for all values of $\mu$, the values of  $S_{bb}$ are more significant than $\log 2$ and hence lie in the fractal region. The values of $S_{bb}$ is much more than $\log 2$. 
  
  Further, we have shown the zoom-in of the Fig. 7(a,b) in the interval (0.23,0.25) in Fig. 8(a,b). For several values of the parameter $\mu$, the values of $S_{b}$ and $S_{bb}$ are computed and plotted using Mathematica. In Fig. 8(a), we observe two peaks at $\mu=0.234$ and $\mu=0.238$. At these values, we also observe the change of libration points from twelve to twenty. It may be the reason behind this. After the value of $\mu=0.238$, we notice a sharp decline in the behaviour of the curve. It implies that the BoA is unpredictable throughout in the range (0.230, 0.238). After that, the BoA lie in non-fractal region. At these two values, we notice similar peaks in the graph of $S_{bb}$ in Fig. 8(b). But in case of boundary basin entropy, we observe that the curve maintains a regular presence in fractal region. The unpredictability of the boundary of BoA is highly unpredictable in this interval. Based on these results, we may think about the possibility of the existence of Wada property. In the next section, we will discuss the existence of the Wada basin boundaries in R6BP.

  \subsection{Existence of Wada basin boundary}

  \begin{figure*}[htb!]
  	\begin{tabular}{ccc}
  		 \includegraphics[scale=.41]{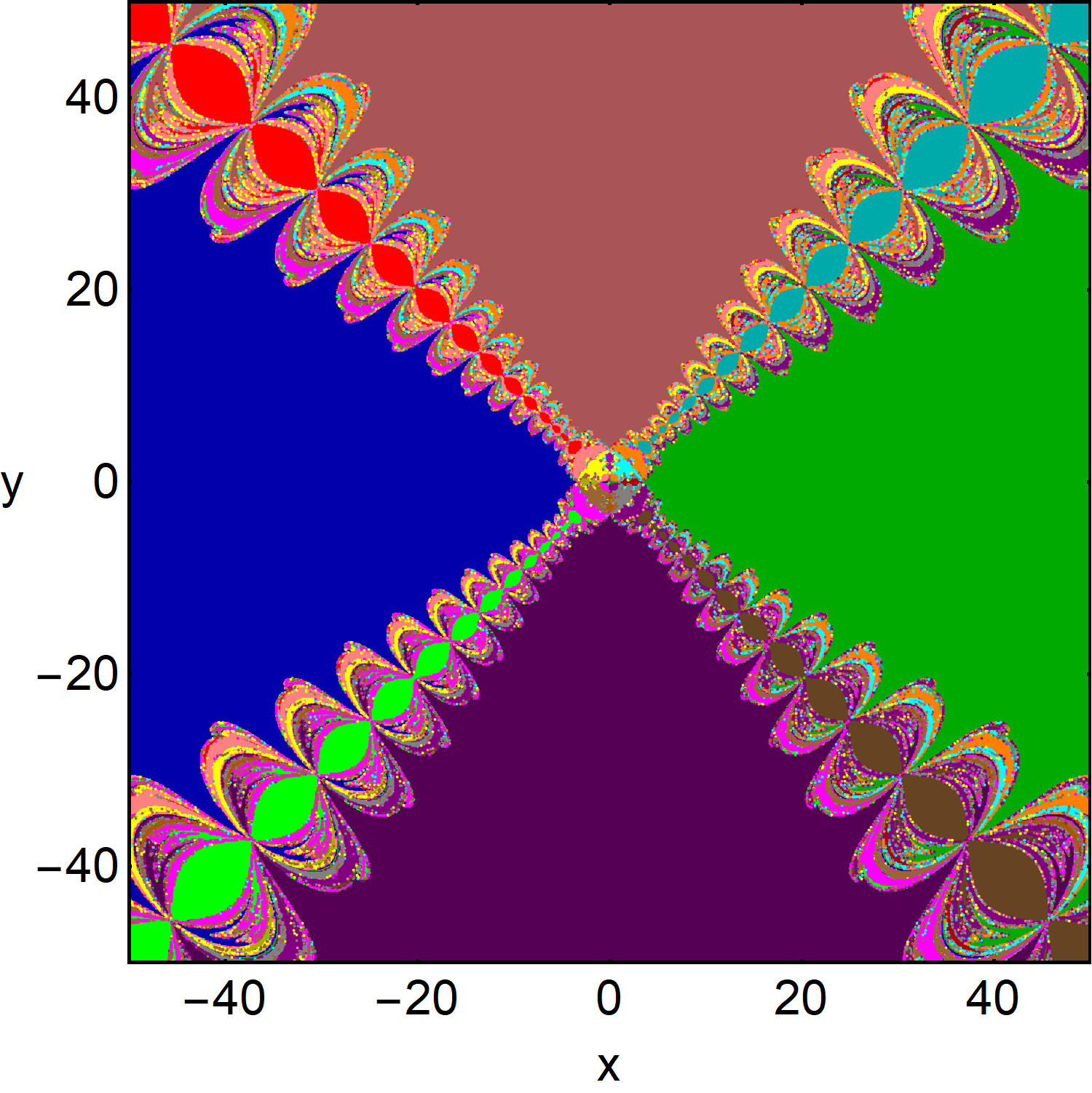}&\includegraphics[scale=.38]{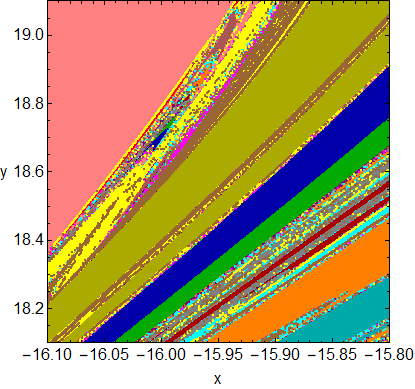} &\includegraphics[scale=.264]{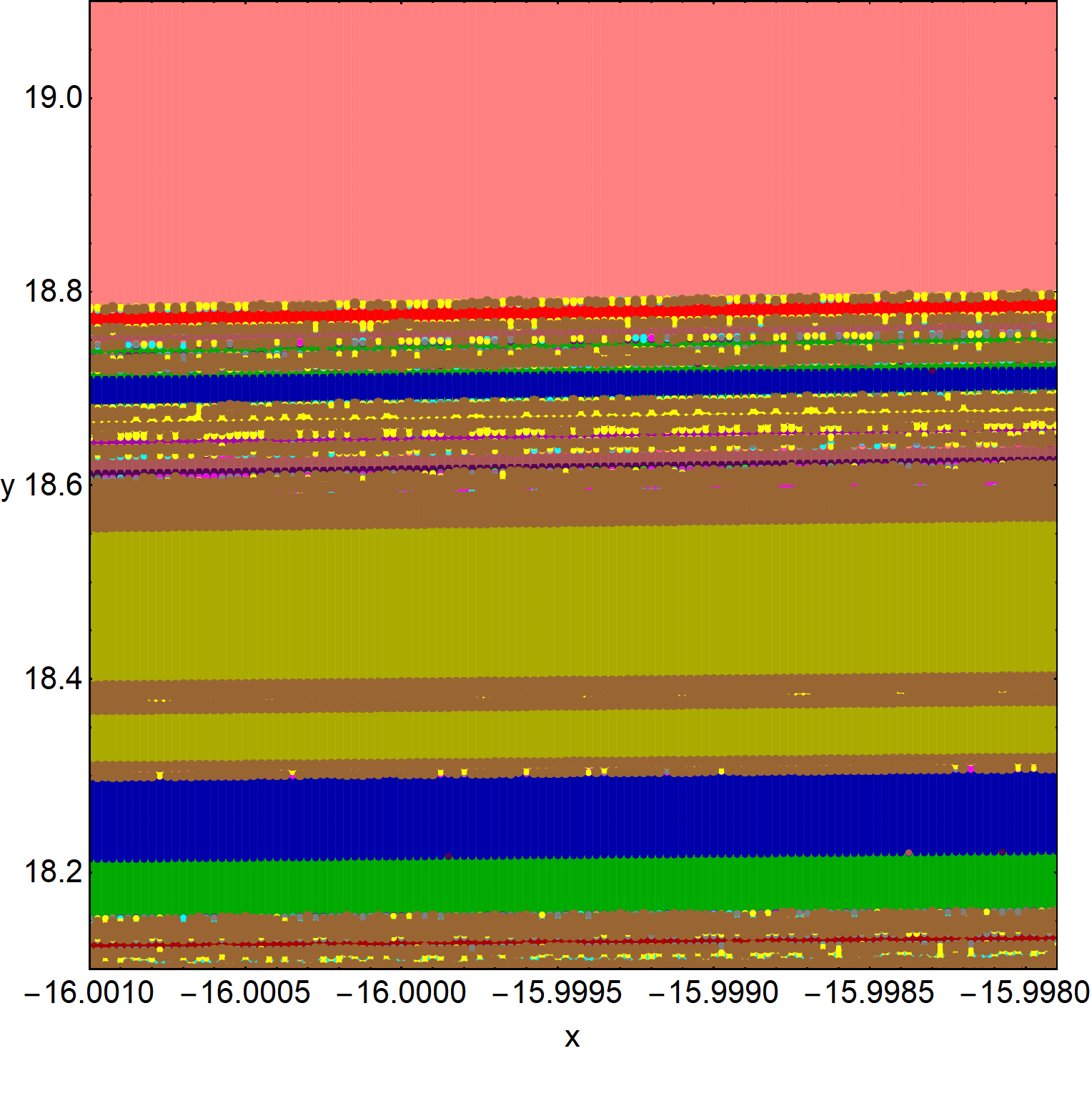}\\
  		(a)  & (b) &(c)
  	\end{tabular}
  	\caption{(a) Graphs of BoA at $\mu$=0.24 by taking larger area on the configuration plane, (b) zoom-in part of the basin boundary (of Fig. (a)) exhibiting the property of Wada for $\mu$=0.24, (c) More zoom-in of the boundary of BoA of Fig. (a).}
  \end{figure*}


\begin{table}[htb!]
	\centering
	\caption{Distribution of the approx 5 million initial conditions towards libration points. This is the case represented in Fig. 6(c) to show the Wada property. Interval on x-axis is (-16.001, 15.998) and y-axis is (18.1, 19.1).}\label{gty}	
	\begin{tabular}{|r|c|l|c|c|c|c|}
		\hline
		\textbf{$\ L_{1}$} & {$ 463238$} & \textbf{$\ L_{8}$} & $ 13921$ & \textbf{$\ L_{15}$} & $ 1609878$\\
		\hline
		\textbf{$\ L_{2}$} & $ 102388$ & \textbf{$\ L_{9}$} & $ 23709$ & \textbf{$\ L_{16}$} & $ 63312 $\\
		\hline
		\textbf{$\ L_{3}$} & $ 572844$ & \textbf{$\ L_{10}$} & $16250$ &\textbf{$\ L_{17}$} & $ 80140$\\
		\hline
		\textbf{$\ L_{4}$} & $ 1164275$ & \textbf{$\ L_{11}$} & $ 119922 $ & \textbf{$\ L_{18}$} & $ 48698$\\
		\hline
		\textbf{$\ L_{5}$} & $ 198809 $ & \textbf{$\ L_{12}$} & $ 30774$ & \textbf{$\ L_{19}$} & $ 119465$\\
		\hline
		\textbf{$\ L_{6}$} & $ 85086$ & \textbf{$\ L_{13} $}& $ 86010$ & \textbf{$\ L_{20}$ }& $ 134252$\\
		\hline
		\textbf{$\ L_{7} $} & $ 82344$ & \textbf{$\ L_{14} $} & {$ 49269$} & {}&{}\\
		\hline
	\end{tabular}
\end{table}

Generally, in a system of two or more variables, there is a possibility of occurrence of a property, known as Wada. A boundary of BoA is said to have the Wada property if for any boundary point p and a smallest positive number $\epsilon>0$, there exist an open neighbourhood centred at 'p' with radius $\epsilon$ containing points of other basins. An obvious implication of this definition is that there is a common boundary of all basins in case of Wada. We also call it a wada basin boundary. Initial conditions lying on these boundaries are highly sensitive, as we can not predict the final destiny of any one of them. Some important articles related to Wada, can be seen in the works of (\cite{Aguirre2009}, \cite{Sanjuan2001}, \cite{Sanjuan2002}, \cite{Daza2015} and \cite{Alvar17a}). The algorithm, to show the existence of Wada property, is not so easy to implement. We have tried to investigate the Wada property in R6BP using the method explained in (\cite{Sanjuan2018}). We have considered the case when $\mu$=0.24. Zoom-in part (in a bigger area in configuration plane)of Fig 2(f) is shown in Fig. 7(a). We observe the presence of all colour in the boundary of BoA. In Fig. 7(b), We have shown the Zoom-in of Fig 7(a). Again we observe the presence of all colour in the boundary. Again, we have increased the resolution of Fig. 7(b) and shown the zoom-in in Fig. 7(c). This time we have considered a very small interval for zoom-in (on $x-axis$ (-16.001, -15.998) and on $y-axis$ (18.1, 19.1)). We have considered approximately five million initial conditions for fine resolution. Again we notice that all colour are present in the boundary shown in Fig. 7(c). We have presented the distribution of five million initial conditions according to their convergence to a particular $L_{i}$ where i=1,2,3,...,20 (Table 5). Thus, we have sufficient evidence to prove the existence of Wada. We have repeated this experiment for all possible values of $\mu$. In all cases,  the evidence of Wada basin boundary is confirmed.

\section{Concluding remarks}
We have considered R6BP with a square configuration for the investigation. The evidence of twelve and twenty libration points for the different values of parameter $\mu$ are crucial enough for further in-depth study. It provides sufficient motivation for the investigation of  N-R BoA.  As the BoA appears to be smooth except boundaries, we have explored the unpredictable region in BoA along the boundaries. The multivariate form of the N-R method and the method of basin entropy is used to study these phenomena. The data obtained from numerical simulations is tabulated in Tables 3 \& 4 and Figs. 4-6, which reflects the originality of the present work.  Tables 3 \& 4 comprises of the details of the convergence of initial conditions towards libration points and time consumed by CPU (in second). The number of iterations needed for the convergence of the initial conditions is shown using Pie-Chart (green colour tone) which is useful, informative and distinct from earlier works. Further, we have established the relationship between the number of iterations needed for convergence and the set of initial conditions on configuration plane $(x, y)$.  The unpredictable nature of boundaries is explored using the concept of basin entropy. Due to the variation of the parameter $\mu$, the effect on the degree of unpredictability of the boundaries of BoA are shown in Figs. 7. The presence of the Wada property is numerically investigated. To the best of our knowledge, this is the first time that such a comprehensive and organised numerical exploration, concerning the BoA, the existence of unpredictability in BoA and the evidence of Wada basin boundary takes place in the R6BP, which is precisely the novelty as well as the significance of the present work.

For all numerical simulations, we have used a machine configured with Intel (R) Core (TM)  i7-8550U CPU 1.80 GHz. In all cases, the computational time of the CPU is also recorded. The main results of numerical simulations can be summarised as follows:
\begin{itemize}
	
	\item The programming for the computation of BoA, basin entropy and plotting of all graphs are done in Mathematica. The execution time taken by CPU for BoA is less than 75 minutes and the classification of initial conditions in this model is given in Table 3 \& 4. As per data available from earlier works, the time specified here is considerably less.
	
	\item The area enclosed by BoA corresponding to all libration points $L_{i}$, i=1, 2, 3... extend to infinity in all cases (Figs. 4(a-f)). We have verified this result by taking extended domain in the configuration plane $(x, y)$ for each case. Also, we find symmetry in BoA about $x$-axis and $y$- axis both.
	
	\item In all cases, the configuration plane $(x, y)$ is a complicated mixture of BoA and extremely fractal regions (Figs. 4(a-f)). These regions are mainly located along the boundaries of BoA. The BoA are intertwined to each other (wherever their combination exists). In these areas, it is not possible to predict the final state of the initial conditions. The unpredictability of the BoA is due to the non-linearity of expressions of the potential responsible for the motion of infinitesimal mass $\acute{m}$ in the presence of the primaries ($m_{1}, m_{2}, m_{3}\ \text{and}\ m_{4}$).  These regions are mainly located in the vicinity of boundaries of BoA.
	
	\item It is found that there are three concentric circles on which the twelve libration points lie and in case of $\mu$=0.24 there are five concentric circles on which twenty libration points lie. It is found that the libration points lie on the same circle attracts approximately the same number of initial conditions. Due to increase in $\mu$, there is a contraction in the domain of convergence of $L_{i}$, i=1, 2, 3,..., 8 whereas the domain of convergence of  $L_{i}$, i=9, 10, 11, 12 expand (Tables 3 \& 4).
	
	\item Based on simulations, we observe that all the initial conditions (one million approximately) on a uniform grid of 1024$\times$1024 converge to the twelve and twenty libration points of the dynamical system with an accuracy of order $10^{-15}$. We do not find any non-converging initial condition. (Table 3 \& 4)
	
	\item  If we look at the nature of BoA, for all cases, $S_{b}$ is less then $\log2$ except for $\mu = 0.22$ and $\mu = 0.23$ (at these values we have computed). At these values, the BoA is not smooth, and the degree of unpredictability confirm the existence of fractal (Fig. 7(a)). However, if we see Fig. 7(b), we find that the degree of unpredictability is on the higher side for all values of $\mu$. Also, due to an increase in $\mu$, there is an increase in the degree of fractality and it slows down after $\mu$=0.22. At $\mu$=0.22, there is higher degree of fractality.
	
	\item Whenever there is change in the values of libration points due to change in the  values of the parameter $\mu$, we observe peak in the graph of $S_{b}$ and $S_{bb}$. These values are 0.234 and 0.238. It is important ot note that after $\mu= 0.238$, BoA is no more unpredictable (Fig. 8(a)). But the boundaries of BoA remains unpredictable throughout the range of $\mu$.   
	
	\item Pie-Charts are introduced to explain the relation between the initial conditions and the number of required iterations needed to achieve the desired level of accuracy, which are shown in Figs. 5(a-f). Also, we can determine the percentage area of initial conditions which converge after 'N' iterations. Also, by simple observation of this chart, one can answer that after which iteration, the maximum number of initial conditions have converged. In Figs. 5(a-f), we notice that 95\% of initial conditions (in approximately one million) converge after 25-35 N-R iterations.
	
	\item  In Fig. 6, we have displayed the graph in blue colour tone. The higher intensity of blue colour tone is allotted to the initial conditions which need more number of N-R iterations to converge. The Figs. 6 (a-f) indicates that the initial conditions lying along the boundaries need more number of iterations to converge. Also, the value of $S_{bb}$ is relatively high along the boundaries (Fig 7(b)). The initial conditions away from the boundaries need less number of iterations to converge. The value of $S_{b}$ is less than $\log 2$ for almost all values of the parameter $\mu$ (Fig. 7(a)). Here we can infer that initial conditions taking more number of iterations to converge are contributing much to the degree of fractality.

	\item The time taken by CPU for the BoA is less than 1 hours for each case (Tables 3 \& 4). Although in these works (\cite{Suraj2019a}, \cite{Suraj2020} and \cite{Suraj2019}), the authors have not given the time taken by CPU for the graph of basins of convergence, we have mentioned it for further comparison. However, for classification of one million initial conditions, CPU time is less than 30 seconds which is relatively less than earlier works.
	
	\item As we notice that the degree of fractality is quite high along the boundaries, we verified the presence of an important property known as 'Wada'. The method adopted for the verification of Wada property is taken from the work of \cite{Sanjuan2013}. At $\mu$=0.24. After taking the sufficient resolution (considering the limit of computation) along the boundaries, we find the presence of all colours in the boundary. The data related to the presence of all colours are given in Table 5. This implies the confirmation of Wada. Thus the existence of Wada basin boundaries is confirmed in the case of R6BP. It shows the novelty of this work.
	
	\item Our results indicate that the effect of the parameter $\mu$ is significant. Due to the presence of $\mu$, we see a noticeable change in the shape of BoA, the existence of fractal and the existence of the Wada property.
\end{itemize}

Based on different outcomes obtained after a systematic and detailed investigation, we can say that the parameter $\mu$ has a considerable impact on the geometry of the BoA. As we have not taken any perturbation in this model, the scholars have a broader scope to work in this model taking 'n' number of perturbations. It is also beneficial for the scholars working in the area of \emph{Celestial Mechanics} as the R6BP has been recently introduced. On the other hand, the proposed result also suggests that the phase space of this model will be exciting and needs an in-depth study. In future, we will try to explore the phase space structure of this model. Further, we will try to study the geometry of the domain of convergence in three-dimensional space as well as the possibility of the properties like Wada basin, riddled basin and many more.


\subsection*{References}

\begin{thebibliography}{0}

	\bibitem[Aguirre(2009)]{Aguirre2009}Aguirre, J., Viana, Ricardo L., Sanjuán, Miguel A. F., 2009. On the fractal structures in nonlinear dynamics, Rev. Mod. Phys., 81, 333--386.

	\bibitem[Sanjuan (2001)]{Sanjuan2001} Aguirre, J., Vallejo, J.C., Sanjuán, M.A.F., 2001. Wada basins and chaotic invariant sets in the Hénon-Heiles system, Phys. Rev. E, 64, 06620.

	 \bibitem[Sanjuan (2002)]{Sanjuan2002} Aguirre, J., Sanjuán, M.A.F., 2002. Unpredictable behavior in the Duffing oscillator: Wada basins. Physica D, 171, 41–51.

    \bibitem[Arribas(2016)]{Arribas16}
    Arribas, M., Abad, A., Elipe, A., Palacios, M., 2016. Out-of-plane equilibria in the symmetric collinear restricted four-body problem with radiation pressure, Astron. Astrophysics., 361: 270.

   \bibitem[Baltagiannis(2011)]{Balta2011}
     Baltagiannis, A.N., Papadakis, K.E., 2011. Equilibrium points and their stability in the restricted four-body problem. International Journal of Bifurcations and Chaos, 21 (8), 2179-2193.

    \bibitem[Celli(2007)]{Celli2007}
     Celli, M., 2007. The central configurations of four masses x, -x, y, -y. Journal of Differential Equations, 235, 668-682.

	\bibitem[Alvar(2017)]{Alvar17} Daza, A., Bertrand, G., Guéry-Odelin, D., Wagemakers, A., Sanjuán, Miguel A. F., 2017. Chaotic dynamics and fractal structures in experiments with cold atoms. Phys. Rev. A 95, 013629.

	\bibitem[Alvar(2016)]{Alvar16} Daza, A., Wagemakers, A., Georgeot, B., Guéry-Odelin, D., Sanjuán, M.A., 2016. Basin entropy: A new tool to analyze uncertainty in dynamical systems, Scient. Rep. \textbf{6}, 31416.

    \bibitem[Sanjuan (2015)]{Daza2015} Daza, A., Wagemakers, A., Sanjuán, Miguel A.F. and Yorke, J.A., 2015. Testing for Basins of Wada. Sci Rep 5, 16579. https://doi.org/10.1038/srep16579.

	\bibitem[Alvar(2017)]{Alvar17a} Daza, A., Wagemakers, A., Sanjuán, M.A.F., 2018. Ascertaining when a basin is Wada: the merging method. Sci Rep 8, 9954. https://doi.org/10.1038/s41598-018-28119-0.

\bibitem[Idrisi(2020)]{Idrisi2020}
Idrisi, M. Javed, Ullah, M. Shahbaz, 2020. Central-body square configuration of restricted six-body problem. New Astronomy, 80, 101381.

\bibitem[Sanjuan (2018)]{Sanjuan2018} Juan, D., Bernal, Jesús, M.S, Miguel Sanjuán, A.F., 2018. Uncertainty dimension and basin entropy in relativistic chaotic scattering, Phys. Rev. E 97, 042214.

\bibitem[Kalvouridis(1999)]{Kalv1999}
Kalvouridis, T.J., 1999. A planar case of the n + 1 body problem: The 'Ring' problem. Astron. Astrophysics. 260, 309-325.

\bibitem[Marchesin(2017)]{Marchesin2017}
Marchesin, M., 2017. Stability of a rhomboidal configuration with a central body. Astrophysics and Space Science, 362, 1-13.

\bibitem[Michalodimitrakis(1981)]{Michalo1981a}
Michalodimitrakis, M., 1981. The circular restricted four-body problem. Astron. Astrophysics. 75, 289-305.

	\bibitem[Dubeibe(2020)]{Dubeibe2020} Osorio-Vargas, J.E., Guillermo A., González., Dubeibe, F.L., 2020. Equilibrium points and basins of convergence in the triangular restricted four-body problem with a radiating body, Int. J. Bifurcation and chaos, 30, 2030003.	

\bibitem[Papadouris(2013)]{Papad2013}
Papadouris, J.P., Papadakis, K.E., 2013. Equilibrium points in the photogravitational restricted four-body problem. Astron. Astrophysics. 344, 21-38.

	 \bibitem[Sanjuan(2013)]{Sanjuan2013} Seoane, J.M., Sanjuán, M.A.F., 2013. On the New developments in classical chaotic scattering, Rep Prog Phys, 76, 016001.

\bibitem[Shoaib(2011)]{Shoaib2011}
Shoaib, M., Faye, I., 2011. Collinear equilibrium solutions of four-body problem. Astrophysics and Astronomy, 32, 411-423.

	\bibitem[Sprott et al.(2015)]{sprott2015} Sprott, J.C., Xiong,  A.,  2015. Classifying and quantifying basins of attraction. Chaos: An Interdisciplinary Journal of Nonlinear Science. 25, 8.

   \bibitem[Suraj(2019)]{Suraj2019c}
   Suraj, M.S., Abouelmagd, E.I., Aggarwal, R., Mittal, A., 2019. The analysis of restricted five-body problem within frame of variable mass. New Astronomy, 70, 12-21.

  \bibitem[Suraj(2019)]{Suraj2019a} Suraj, M.S., Aggarwal, R., Mittal, A. Asique, M. C., Sachan, P., 2019. On the perturbed photogravitational restricted five-body problem: the analysis of fractal basins of convergence. Astron. Astrophysics. 364, 87. https://doi.org/10.1007/s10509-019-3575-3.

   \bibitem[Suraj(2019)]{Suraj2019b} Suraj, M.S., Sachan, P., Zotos, E.E., Mittal, A., Aggarwal, R., 2019 On the Newton-Raphson basins of convergence associated with the libration points in the axisymmetric restricted five-body problem: the concave configuration. Int. J. Non-Linear Mech. 112, 25–47.

   \bibitem[Suraj(2020)]{Suraj2020} Suraj, M.S., Mittal, A, Kaur, C, Aggarwal, R.,  2020. Analysis of Copenhagen problem with a repulsive quasi‐homogeneous Manev‐type potential within the frame of variable mass., Astron. Nachr.  1– 14, https://doi.org/10.1002/asna.202013640.	  	

	\bibitem[Suraj(2019)]{Suraj2019}Suraj, M.S., Sachan, P., Zotos, E.E.,  Mittal, A and Aggarwal, R., 2019. On the fractal basins of convergence of the libration points in the axis-symmetric five-body problem: The convex configuration, Int. J. of Non-Linear Mech., 109, 80-106.

 \bibitem[Wolfram (2017)]{wolf2017} Wolfram Research, Inc. Mathematica Version 11.0.1. Champaign, IL, (2017)
    	
\end{thebibliography}
\end{document}